\newif\ifshowcomments
\showcommentstrue 

\documentclass[a4paper,11pt]{article}
\usepackage{jcappub}

\pdfoutput=1

\usepackage{tabularx,booktabs}
\newcolumntype{Y}{>{\centering\arraybackslash}X}

\usepackage[utf8]{inputenc}
\usepackage[font=small,labelfont=bf,tableposition=top]{caption}
\usepackage{float}
\usepackage{adjustbox}
\usepackage{arydshln}

\usepackage{amsmath,amssymb,amsbsy,amstext, amsthm, simplewick, amsfonts, bm}
\usepackage{hyperref}
\usepackage{graphicx}
\usepackage{siunitx}
\usepackage{upgreek}
\usepackage{framed}
\usepackage{wrapfig}
\usepackage{multirow}
\usepackage{subfig}
\usepackage{bbm}
\usepackage[svgnames,dvipsnames,x11names]{xcolor}
\usepackage{enumitem}
\usepackage{fontawesome}

\usepackage[top=2.5cm,bottom=3.cm,right=2.5cm,left=2.5cm]{geometry}

\pdfoutput=1
\usepackage{natbib,ifthen}
\usepackage{enumerate}
\usepackage{empheq}
\usepackage{feynmp}
\usepackage[english]{babel}
\usepackage{amsmath,amssymb,amsbsy,amstext, amsthm, simplewick,stackrel}
\usepackage{hyperref}
\usepackage{graphicx}
\usepackage{amsfonts}
\usepackage{amssymb}
\usepackage{upgreek}
 \usepackage{exscale,relsize}
 \usepackage[makeroom]{cancel}
\usepackage{float}
\usepackage{framed}
\usepackage{array}
\newcolumntype{C}[1]{>{\centering\arraybackslash}p{#1}}

\usepackage{placeins}
\newcommand{\MyFloatBarrier}{\FloatBarrier}  
\usepackage{color}
\usepackage{epsfig}
\usepackage{bm}
\usepackage{wasysym}
\usepackage{mathrsfs}

\usepackage{enumitem}
\usepackage{fontawesome}

\newcommand{\appropto}{\mathrel{\vcenter{
  \offinterlineskip\halign{\hfil$##$\cr
    \propto\cr\noalign{\kern2pt}\sim\cr\noalign{\kern-2pt}}}}}

\newcommand{\comment}[1]{}

\newcommand{\beq}{\begin{equation}}
\newcommand{\eeq}{\end{equation}}
\newcommand{\be}{\begin{equation}}
\newcommand{\ee}{\end{equation}}
\newcommand{\bea}{\begin{eqnarray}}
\newcommand{\eea}{\end{eqnarray}}
\newcommand{\bsp}{\begin{split}}
\newcommand{\esp}{\end{split}}

\newcommand{\hMpc}{\ h^{-1}\text{Mpc}}

\renewcommand{\vec}[1]{\bm{#1}}

\newcommand{\vx}{\vec x}
\newcommand{\vv}{\vec v}
\newcommand{\vs}{\vec s}

\newcommand{\vk} {{\boldsymbol k}}

\newcommand{\vq} {{\boldsymbol q}}
\newcommand{\vz} {{\boldsymbol z}}

\newcommand{\vpsi}{\vec \psi}

\newcommand{\G}{\mathcal{G}}

\newcommand{\ihMpc}{\; h\text{Mpc}^{-1}}

\definecolor{darkgreen}{RGB}{0,120,0}

\newcommand{\secref}[1]{Section \ref{se:#1}}

\newcommand{\eq}[1]{(\ref{eq:#1})} 
\newcommand{\eqq}[1]{Eq.~(\ref{eq:#1})} 
\newcommand{\fig}[1]{Fig.~\ref{fig:#1}} 
 
\newcommand{\app}[1]{Appendix~\ref{app:#1}}

\definecolor{darkgreen}{RGB}{0,120,0}
\ifshowcomments
\newcommand{\MS}[1]{{\color{darkgreen}{MS: #1}}}
\newcommand{\Marko}[1]{\textcolor{blue}{(Marko: #1)}}
\newcommand{\va}[1]{\textcolor{green}{(VA: #1)}}
\newcommand{\misha}[1]{\textcolor{magenta}{(MI: #1)}}
\newcommand{\MZ}[1]{{\color{red}{MZ: #1}}}
\newcommand{\oliver}[1]{\textcolor{purple}{(Oliver: #1)}}
\else
\newcommand{\MS}[1]{}
\newcommand{\Marko}[1]{}
\newcommand{\va}[1]{}
\newcommand{\MZ}[1]{}
\newcommand{\misha}[1]{}
\newcommand{\oliver}[1]{}
\fi

\usepackage{colortbl}
\definecolor{rp}{cmyk}{0.2, 1, 0.6, 0}
\definecolor{green2}{cmyk}{0, 1, 0.5, 0}
\definecolor{lightgreen}{cmyk}{0.2, 0, 0.2, 0.2}
\definecolor{lightgray}{cmyk}{0.1,0.2,0,0.1}
\definecolor{lightgray2}{cmyk}{0.4,0.4,0,0.8}
\definecolor{black}{cmyk}{1.0,1.0,1.0,1.0}

\definecolor{lightgreen}{cmyk}{0.2, 0, 0.2, 0.2}
\definecolor{lightgray}{cmyk}{0.1,0.2,0,0.1}
\definecolor{lightgray2}{cmyk}{0.1,0.1,0,0.1}

\makeatletter
\newlength{\apb@width}
\newcommand{\autoparbox}[2][c]{\settowidth{\apb@width}{#2}\parbox[#1]{\apb@width}{#2}}

\makeatother

\def\beq{\begin{equation}}
\def\eeq{\end{equation}}

\def\bea{\begin{eqnarray}}
\def\eea{\end{eqnarray}}

\def\beq{\begin{equation}}
\def\eeq{\end{equation}}
\def\bea{\begin{eqnarray}}
\def\eea{\end{eqnarray}}

\def\G{{\cal G}}

\def\0{{\boldsymbol 0}}
\def\k{{\boldsymbol{k}}}
\def\q{{\boldsymbol{q}}}

\def\z{{\boldsymbol{z}}}

\DeclareRobustCommand{\SkipTocEntry}[4]{}

\title{Modeling Galaxies in Redshift Space at the Field Level}

\author[a]{Marcel Schmittfull
 \footnote{\texttt{mschmittfull@gmail.com}}
}
\author[b]{Marko Simonovi\'c
 \footnote{\texttt{marko.simonovic@cern.ch}}
} 
\author[c,d]{Mikhail M.~Ivanov
}
\author[a,e]{Oliver H.\,E.~Philcox
}
\author[a]{Matias Zaldarriaga
}

\affiliation[a]{School of Natural Sciences, Institute for Advanced Study,\\1 Einstein Drive, Princeton, NJ 08540, USA}
\affiliation[b]{Theoretical Physics Department, CERN,\\1 Esplanade des Particules, Geneva 23, CH-1211, Switzerland}
\affiliation[c]{Center for Cosmology and Particle Physics, Department of Physics,
New York University,\\
New York, NY 10003, USA}      
\affiliation[d]{Institute for Nuclear Research of the
Russian Academy of Sciences, \\ 
60th October Anniversary Prospect, 7a, 117312
Moscow, Russia}
\affiliation[e]{Department of Astrophysical Sciences, Princeton University,\\ Princeton, NJ 08540, USA}%

\date{\today}

\abstract{We develop an analytical forward model based on perturbation theory to predict the redshift-space galaxy overdensity at the field level given a realization of the initial conditions. 
We find that the residual noise between the model and simulated galaxy density has a power spectrum that is white on large scales, with size comparable to the shot noise. 
In the mildly nonlinear regime, we see a $k^2\mu^2$ correction to the noise power spectrum, corresponding to larger noise along the line of sight and on smaller scales.
The parametric form of this correction has been predicted on theoretical grounds before, and our simulations provide important confirmation of its presence. 
We have also modeled the galaxy velocity at the field-level and compared it against simulated galaxy velocities, finding that about $10\%$ of the galaxies are responsible for half of the rms velocity residual for our simulated galaxy sample.
}

\begin{document}

\maketitle

\section{Introduction}

The clustering of galaxies is an increasingly important cosmological probe of the low-redshift Universe. 
Building on the success of recent surveys like SDSS BOSS/eBOSS \cite{Alam:2016hwk,Alam:2020sor}, multiple experiments will increase the volume and the number of observed galaxies in the near future, including DESI \cite{Aghamousa:2016zmz}, Subaru HSC/PFS \cite{2014PASJ...66R...1T,Hikage:2018qbn}, Euclid \cite{Amendola:2016saw}, Vera Rubin Obervatory/LSST \cite{Abell:2009aa}, SPHEREx \cite{Dore:2014cca}, Roman Telescope/WFIRST \cite{2019arXiv190205569A}, and others.
Having lower cosmic variance and shot noise, these experiments demand accurate theoretical modeling at the percent-level or better to allow for unbiased cosmological parameter inference (e.g., \cite{Nishimichi:2020tvu}).
Numerical simulations can provide such modeling for the clustering of dark matter, and while they are numerically expensive, interpolation schemes may be employed to overcome the computational challenge of this approach.
However, the clustering of galaxies is biased with respect to that of the dark matter, in an a priori unknown way that depends on the type of galaxies observed, for example on their mass or the environment in which they formed. 
While numerical schemes can be implemented to place galaxies in collapsed dark matter halos in simulations, it is not known what family of assignment schemes is correct for a given galaxy sample, and how to make this fully general is a topic of ongoing research; see e.g.~\cite{Hadzhiyska:2019xnf,Hadzhiyska:2020iri} for two recent studies.
\vskip 4pt

An alternative approach, which we follow here, is to describe the galaxy density field perturbatively, including all possible terms in the bias expansion allowed by the symmetries of the problem~\cite{Desjacques:2016bnm}.
In this case, one has an analytical model for the clustering of galaxies in redshift space, which can readily be used for cosmological parameter inference using standard MCMC techniques~\cite{Nishimichi:2020tvu,Chudaykin:2020aoj,DAmico:2020kxu}.
Apart from gravitational and biasing nonlinearities, one other important part of this model is stochastic noise. 
Using the correct parametric form of the power spectrum of this noise is critical for cosmological parameter analyses from galaxy surveys, because the noise power spectrum contributes to the total model power spectrum so that any missing ingredient in the noise model could lead to biases in cosmological parameter inferences.
\vskip 4pt 

The simplest approach is to assume a white noise power spectrum, roughly of the size of the shot noise.
However, it can be argued based on symmetries that the stochastic noise of the galaxy overdensity in redshift space should have corrections on the white power spectrum, scaling as $k^2$ and $k^2\mu^2$, where $\mu$ is the cosine with respect to the line of sight and $k$ is the wavenumber of the galaxy overdensity~\cite{Perko:2016puo}.
This prediction can be tested using simulated galaxy catalogs.
When sample variance is present and a simulated galaxy power spectrum is fitted with multiple parameters  (e.g., galaxy bias parameters, counter terms, and stochasticity parameters), it is challenging to identify the exact parametric form of the noise power spectrum.
A more powerful technique to characterize the form of the noise is to measure it directly, subtracting the model prediction from the simulated galaxy overdensity field. 
This avoids sample variance if model prediction and simulation are computed for the same initial random seed.
While this noise has been measured for simulations of dark matter \cite{Baldauf:2015zga,Taruya:2018jtk}, dark matter halos in real space \cite{Marcel1811}, and 21cm radiation \cite{McQuinn:2018zwa}, it has not been investigated for halos or galaxies in redshift space. 
This is the subject of this paper.
A primary goal is to use field level methods to conduct a sample-variance-free test of the anisotropic noise prediction. At the same time, such measurement is also a test of the perturbative model for galaxy clustering in redshift space. Such investigations of the noise properties, combined with efforts to model the galaxy power spectrum at the subpercent level,
provide a solid theoretical foundation for analyzing future galaxy surveys.
\vskip 4pt

The paper is outlined as follows. We first model and measure the velocities of a simulated sample of galaxies.
Based on the redshift space distortions generated by these velocities, we introduce a model for the galaxy density in redshift space.
We then measure the error of this model by comparing against simulated galaxies in redshift space.
We characterize its scale dependence and compare it against theoretical expectations, before concluding.
\vskip 4pt

An accompanying Python software package, \textsc{perr}, is available online \href{https://github.com/mschmittfull/perr}{\faGithub} \footnote{\url{https://github.com/mschmittfull/perr}}.
It is based on \textsc{nbodykit} \cite{Hand:2017pqn} \href{https://github.com/bccp/nbodykit}{\faGithub} \footnote{\url{https://github.com/bccp/nbodykit}} and can be used to generate the 3D models for the velocity and galaxy density and compare them with simulations.
\vskip 4pt

\section{Galaxy velocities and redshift space displacements}

We start by discussing the velocity of galaxies in real space, which determines the redshift space displacement, i.e.~the line-of-sight displacement that must be applied to galaxies' real space positions to obtain their observed positions in redshift space. We review a perturbative model for this redshift space displacement and compare it against simulations. Later, we will build on this to obtain a model of the galaxy overdensity in redshift space and test that against simulations.
\vskip 4pt

\subsection{Velocity of Lagrangian particles}

We can model velocities following Matsubara \cite{2008PhRvD..77f3530M} (also see, e.g., \cite{2013MNRAS.429.1674C}).
In redshift space $\vs$, the location of an object at $\vx$ is mis-identified due the peculiar velocity $\vv=a\dot\vx$ along the line of sight  $\hat \vz$,
\begin{align}
  \label{eq:1}
  \vs = \vx + \frac{\hat \vz\cdot \vv(\vx)}{aH}\hat \vz\;,
\end{align}
where $a$ is the scale factor and $H$ Hubble parameter.
Without redshift space distortions (RSD), the relationship between Lagrangian coordinates $\vq$ and Eulerian coordinates $\vx$ is given by the nonlinear displacement $\vpsi$,
\begin{align}
  \label{eq:2}
  \vx = \vq+\vpsi(\vq)\;.
\end{align}
Including RSD, this becomes
\begin{align}
  \label{eq:3}
  \vs = \vq+\vpsi(\vq) + \frac{\hat \vz\cdot \vv(\vx)}{aH}\hat \vz.
\end{align}
In perturbation theory, the velocity field is related to the time derivative of the displacement field 
\begin{align}
  \label{eq:EulVel}
  \vv(\vx) &= a\dot\vx =a\dot\vpsi=a\sum_{n=1}^\infty nfH\vpsi_n(\vq)\;,
\end{align}
where we have used the perturbative expansion $\vpsi=\sum_n\vpsi_n$. Note that $\vpsi_n\propto D^n(z)$, so that $\dot\vpsi_n=nfH\vpsi_n$, where $D(z)$ is the linear growth factor and $f\equiv d \log D/ d \log a$.
Therefore, we can write
\begin{align}
  \label{eq:5}
  \vs = \vq+\vpsi^{s}(\vq) \;.
\end{align}
The redshift space displacement $\vpsi^{s}(\vq)$ can be written in compact form as follows
\begin{align}
  \label{eq:6}
  \vpsi^{s}(\vq) = \sum_{n=1}^\infty R^{[n]} (\hat \z) \cdot \vpsi_n (\q) \; ,
\end{align}
where the matrices $R^{[n]} (\hat\z)$ are defined as
\be
R^{[n]}_{ij} (\hat\z) \equiv \delta_{ij} + nf\hat z_i \hat z_j \;.
\ee
This standard result describes the mapping of Lagrangian particles at $\vq$ to their Eulerian redshift-space coordinates $\vs$, using the velocity predicted by Lagrangian perturbation theory.
\vskip 4pt

\subsection{Continuous velocity field}
Before using the above mapping of Lagrangian particles to Eulerian redshift space, let us go one step back and consider the velocity field itself, so we can compare it between the model and simulations.
To compute the velocity as a continuous field in Eulerian space one can proceed as follows.
First, the continuity equation in Eulerian space gives
 \begin{align}
 \label{eq:Continuity}
    a\dot\delta(\vx) + \nabla\cdot[(1+\delta(\vx))\vv(\vx)] = 0 \;.
\end{align}
Second, the Lagrangian Perturbation Theory (LPT) expression for the time derivative of the Eulerian density in Fourier space is
\begin{align}
\label{eq:LPTDeltaDot}
    a\dot\delta(\vk)
   &= a\frac{\partial}{\partial t}\int d^3\vq\, e^{i\vk\cdot(\vq+\vpsi(\vq,t))}
    = ai\vk\cdot\tilde{\dot\vpsi}(\vk)\;.
\end{align}
Here, we defined the real-space shifted field
\begin{align}
    \tilde{\dot\vpsi}(\vk) \equiv  \int d^3\vq \,\dot\vpsi(\vq) e^{i\vk\cdot(\vq+\vpsi(\vq))}\;.
\end{align}
This is similar to the Zel'dovich approximation, moving Lagrangian particles from $\vq$ to $\vq+\vpsi$, but weighing each particle by $\dot\vpsi$; this operation is analogous to the shifted bias operators in \cite{Marcel1811}.
Combining Eqs.~\eq{Continuity} and \eq{LPTDeltaDot} shows that the curl-free part of the momentum density is 
\begin{align}
  (1+\delta(\vx))\vv(\vx)=-a\tilde{\dot\vpsi}(\vx) \;.
\end{align}
The resulting curl-free velocity or RSD displacement is given by
\begin{align}
\label{eq:vmodel}
\frac{\vv(\vx)}{aH}=-f\sum_{n=1}^\infty \frac{n\widetilde{\vpsi_n}(\vx)}{1+\delta(\vx)}
\equiv -f\sum_{n=1}^\infty n\widehat{\vpsi_n}(\vx)
\;.
\end{align}
Note that we have used
\begin{align}
    \dot\vpsi(\vq) = \sum_{n=1}^\infty nfH \vpsi_n(\vq)\;.
\end{align}
The equations above are exact if $\vpsi$ is the true displacement field. 
In practice, we use the linear displacement for shifting, and we truncate the sum over $n$ keeping only $n=1$.
\vskip 4pt

\subsection{Evaluating the continuous velocity model in a 3D box}

To obtain the shifted $n$th order displacement $\widetilde{\vpsi_n}(\vx)$ in \eqq{vmodel}, Lagrangian particles are weighted by $\vpsi_n(\vq)$ and then shifted from their Lagrangian position $\vq$ by the real-space displacement $\vpsi_1(\vq)$. When painting the particles to a regular grid, particles are summed up, including their weights.
This is the same procedure as for the shifted bias operators defined in \cite{Marcel1811}.
Instead of dividing by $1+\delta$ in \eqq{vmodel}, one can use a modified painting scheme that divides by the number of particles contributing to each cell.
This is denoted with a hat on the right-hand-side of \eqq{vmodel}, and this is what we will use in the following. 
In detail, we implement the modified painting scheme for evaluating the above model for the continuous velocity field in a 3D box as follows.

\begin{itemize}
    \item Place particles on a regular grid in Lagrangian space. Call their positions  $\vq_i$.
    \item Compute $\dot\vpsi_{1,i}=\dot\vpsi_1(\vq_i)=fH \vpsi_1(\vq)$ 
    for each particle.
    \item Shift each particle to $\vx_i=\vq_i+\vpsi_1(\vq_i)$.
    \item Paint the shifted catalog to a grid, weighting each particle by $\dot\vpsi_{1,i}$. 
    For each cell, divide by the number of particles contributing to that cell. 
    \item The resulting field  $\widehat{\dot\vpsi}(\vx)$ as a function of Eulerian coordinates is our model for the Eulerian velocity.
 \end{itemize}
 Due to the averaging operation in the painting step, the field value does not increase if more $\vq$ particles with the same velocity end up in the same region; this ensures that we indeed obtain the velocity and not the momentum field.
The procedure is similar to the one used to generate shifted bias operators \cite{Marcel1811}, except that we are now shifting $\dot\vpsi(\vq_i)$ instead of bias operators $\mathcal{O}(\vq_i)$, and we modify the painting scheme such that particles contributing to a cell are averaged instead of summed. 
\vskip 4pt

\begin{figure}[tbp]
\centering
\includegraphics[width=1.0\textwidth]{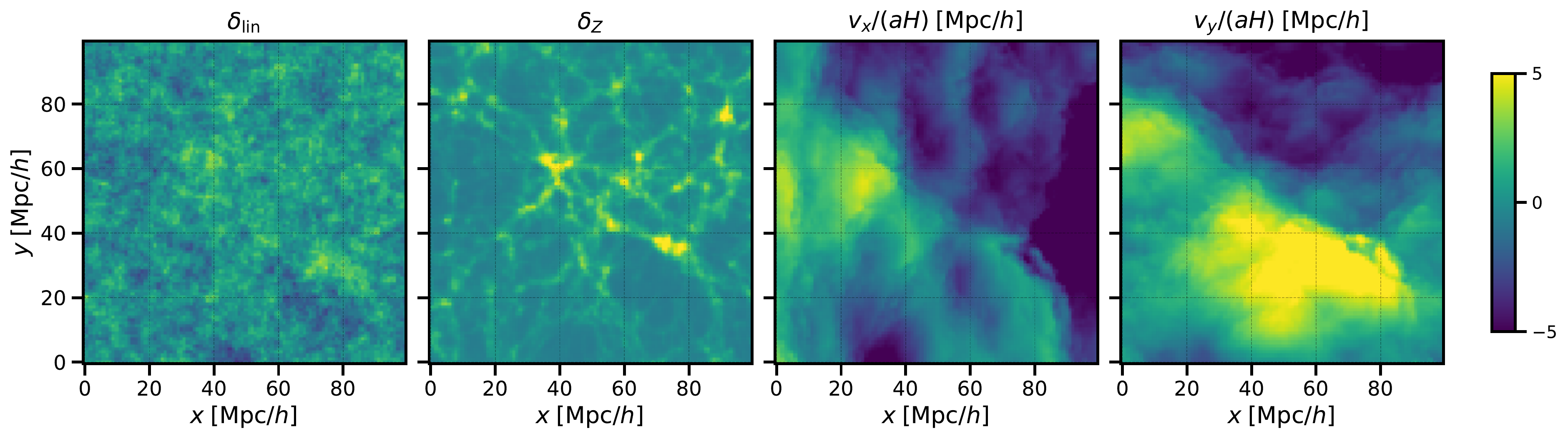}
\caption{2D slice of the linear density, Zel'dovich density, and $x$- and $y$-component of the continuous velocity field predicted by \eqq{vmodel} for $n_\mathrm{max}=1$.
The predicted velocity field is coherent over tens of Megaparsecs, with most regions flowing towards the cluster and filament in the center of the slice.
To generate the Zel'dovich density and the velocity prediction, $1536^3$ particles in a Lagrangian space box with $L=500\hMpc$ were shifted by the first-order displacement. 
All fields are evaluated at redshift $z=0.6$.
}
\label{fig:VelModelSlices}
\end{figure}

A 2D slice of the continuous Eulerian velocity field generated with this method is shown in the two right panels in \fig{VelModelSlices}.
This shows that the model predicts large-scale flows that are coherent over tens of megaparsecs and represent flows towards large-scale overdense regions by a few megaparsecs.
\vskip 4pt

\subsection{Comparison with simulated velocity}
\label{se:VelSimComparison}
\begin{figure}[tbp]
\centering
\includegraphics[width=0.99\textwidth]{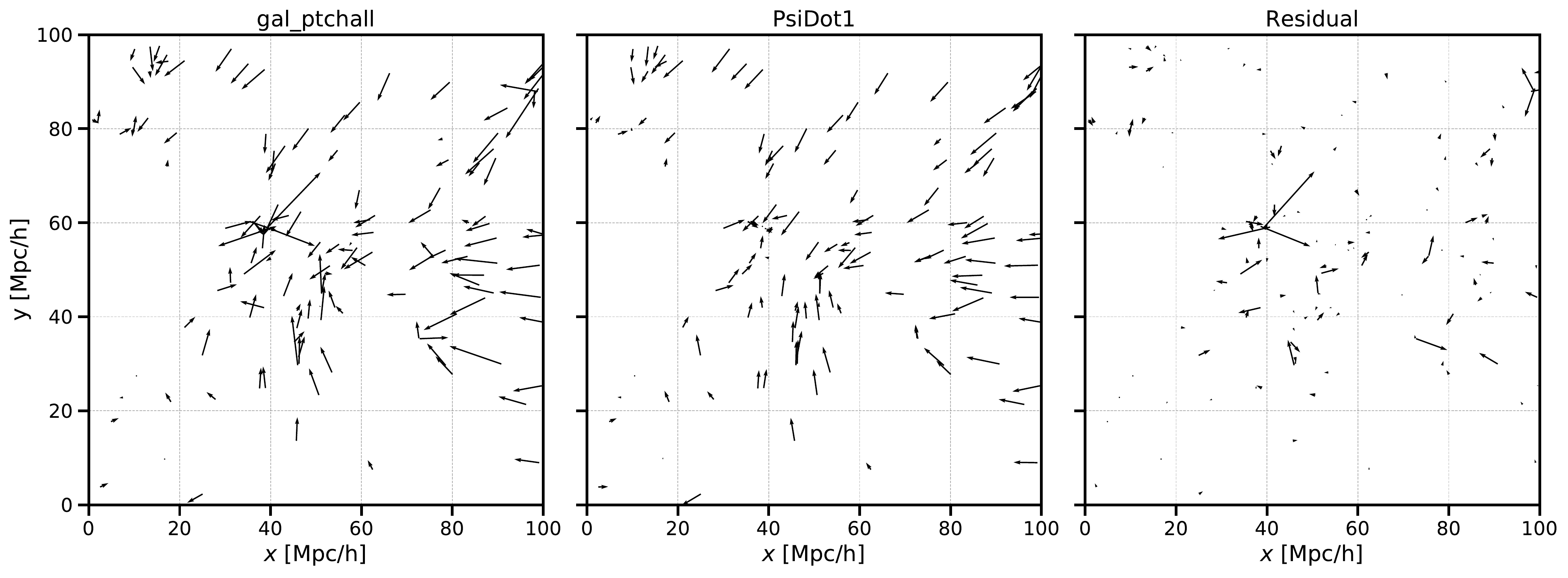}
\caption{RSD displacements in Mpc/h in the $x-y$ plane, for simulated galaxies (left), the model prediction from \eqq{vmodel} evaluated at galaxy positions (center), and their residual (right). The model captures the large-scale bulk flows rather well. The largest mistakes happen in clustered regions where the velocity vector in the simulation is large and goes in random directions at nearly the same location; these are likely satellites with virial velocity.
Since the velocity goes in opposite directions in nearly the same location, there is little hope to model this deterministically, so it should be regarded as an unpredictable noise contribution.
}
\label{fig:PTChallVel}
\end{figure}

To test the above model for the velocity field, we compare it against velocities of different objects measured in an N-body simulation (in real space).
In general it is not known exactly which objects in an N-body simulation correspond to galaxies observed by a specific galaxy survey.
We produce a galaxy sample that approximately reproduces observed properties of SDSS BOSS CMASS galaxies following the procedure of \cite{Nishimichi:2020tvu}.\footnote{The \textsc{Rockstar} \cite{Rockstar}
\href{https://bitbucket.org/gfcstanford/rockstar}{\faBitbucket} phase-space halo finder is used to identify halos and subhalos in the snapshot of a dark-matter-only N-body simulation with $1536^3$ particles in a $L=500\hMpc$ periodic box at redshift $z=0.6$, run with 
\textsc{MP-Gadget} \cite{yu_feng_2018_1451799} \href{https://github.com/MP-Gadget/MP-Gadget}{\faGithub}.
These halos and subhalos are then populated with galaxies with a probability that depends on the virial mass of the object; see Eq.~(1) in \cite{Nishimichi:2020tvu}.
We choose their CMASS1 parameters, i.e.~$\log_{10}M_\mathrm{min}[h^{-1}\text{M}_\odot]=12.97$
and $\sigma_{\log_{10}M}=0.35$.}
Note that this does not explicitly populate halos with centrals and satellites using an halo occupation distribution (HOD), but uses subhalos found with the phase-space halo finder \textsf{Rockstar} \cite{Rockstar} and selects them with a soft mass cut to represent galaxies.
This accounts for the velocity offsets between the halo center of mass and central subhalos \cite{Rockstar}, where galaxies are expected to form. Also, satellites are based on actual subhalo positions and velocities within larger halos rather than assigning satellites with some manual prescription (e.g., using the velocity of random dark matter particles in a halo or an NFW profile).
The velocities of these mock galaxies, converted to the corresponding RSD displacement, are shown in the left panel of \fig{PTChallVel}.
The middle panel shows the analytical prediction using the model from the last section, reading out the continuous velocity field at the location of the simulated galaxies, and the right panel shows the residual displacement between simulation and model.
\vskip 4pt

This shows that the analytical model describes the large-scale velocity field of galaxies rather well.
But it fails in some highly clustered regions, where the residual displacement can be tens of Mpc and is pointing in random directions.
These galaxies are subhalos whose velocity is close to the virial velocity of their parent halo rather than the large-scale velocity field in their vicinity.
Given the random direction of these velocity offsets, it seems challenging to model these velocities with any deterministic model.
For the galaxies in redshift space, this can be regarded as the well-known Fingers of God effect, which corresponds to random motions of satellite galaxies along the line of sight, leading to a relative suppression of clustering along the line of sight.
\vskip 4pt

\begin{figure}[tbp]
\centering
 \includegraphics[width=0.45\textwidth]{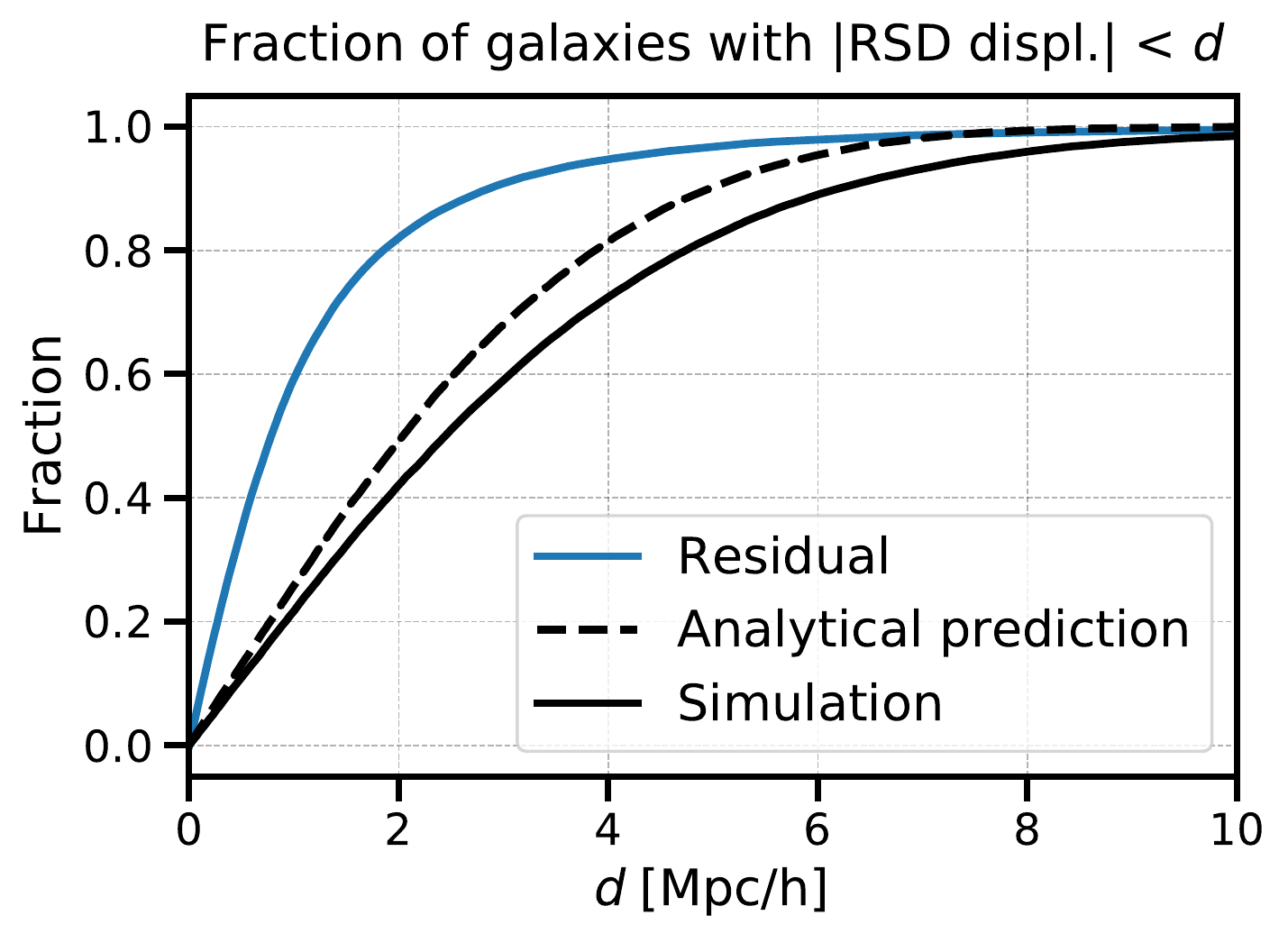}
\includegraphics[width=0.45\textwidth]{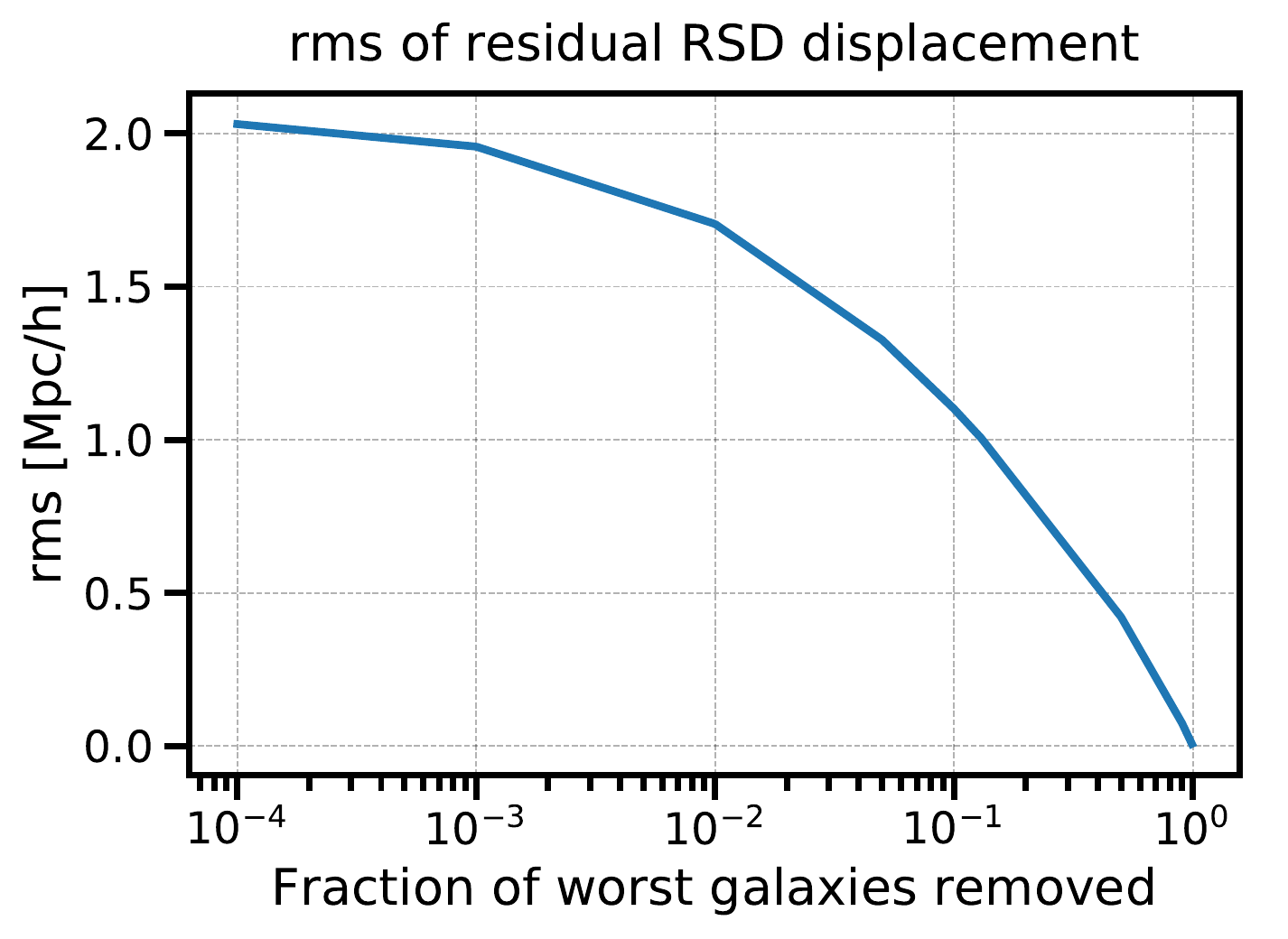}
\caption{\emph{Left panel}: Fraction of galaxies for which the residual RSD displacement (PT challenge galaxies minus shifted $\dot\psi$ model) is smaller than some value $d$.
\emph{Right panel}:  Rms residual RSD displacement after removing different fractions of galaxies with the worst residual.
Removing the worst 13\% of galaxies reduces the rms residual displacement by a factor of 2.
}
\label{fig:fractions}
\end{figure}

From \fig{PTChallVel} it is clear that only a minority of galaxies have these large velocities and corresponding large RSD displacements which are very discrepant with the analytical model prediction, while the prediction is rather accurate for the majority of galaxies.
Indeed, we find that the model predicts the simulated RSD displacement with an error of less than $2\hMpc$ for 80\% of the galaxies and with an error less than $3\hMpc$ for 90\% of the galaxies; see the left panel of \fig{fractions}. This means that only 10-20\% of galaxies have velocities that are crudely wrong.
Removing such galaxies could enable perturbation theory to reach smaller scales when modeling the power spectrum in redshift space, at the expense of a small increase in shot noise.
Removing the 13\% worst-modeled galaxies reduces the rms displacement error by a factor of 2, from $2\hMpc$ to $1\hMpc$.
In other words, about half the rms displacement error ($1\hMpc$) comes from the worst 13\% of galaxies (these have a residual displacement $>2.5\hMpc$), while the other half of the rms ($1\hMpc$) comes from the best 87\% of galaxies (these have residual displacement $<2.5\hMpc$).\footnote{For comparison, the rms RSD displacement is $3.9\hMpc$ for the simulated galaxies and $3.0\hMpc$ for the model prediction evaluated at galaxy positions.}
Identifying these satellite galaxies observationally in redshift space is a challenging task; though see e.g.~\cite{Rodriguez:2020fos} for recent progress and application to SDSS data.\footnote{Typical approaches seek to identify clustered groups in the observed data. Nonlocal selections like this can induce additional scale dependence in the power spectrum.
Another approach could be to use summary statistics in Fourier space. For example, one could optimize weights of galaxies such that the small-scale power spectrum quadrupole is maximized, corresponding to a suppression of the Fingers of God, while keeping the small-scale power spectrum monopole (shot noise) small to ensure that most galaxies are still included. 
A concrete way would be to maximize the ratio of the power spectrum quadrupole over monopole at high $k$.
The optimal weights can be found by solving an eigenvalue problem. 
Going further, one can impose a prior on the weights, e.g.~to set some fraction of them to 0 and the rest to 1, or to follow the halo mass function to implement halo mass weighting and suppress shot noise.
To enforce non-negative weights one can use a sigmoid function.
The resulting cost function can be optimized using gradient descent, noting that derivatives of power spectrum multipoles with respect to galaxy weights can be evaluated using FFTs.
As long as the weights are parameterized as functions of observed galaxy properties, the resulting weighted galaxy sample is still a biased tracer and can be described by the same galaxy bias model as in conventional galaxy clustering analyses.
We leave it to future work to explore this idea.}
\vskip 4pt

Of course the results above are specific to the mock galaxies generated, and they will differ for other tracers and models. We briefly discuss this in \app{OtherVelocityPlots}.
We also note that it is challenging to quantify the velocity error in more detail, for example using power spectra, because it is difficult to compute a continuous velocity field from a discrete tracer as it is unclear how to define the velocity field at locations with no objects.
While we will use the model prediction for the velocity field to model the galaxy clustering in redshift space, it may also be useful in its own right for modeling the kSZ effect or other cosmological probes of the velocity field.
\vskip 4pt

\MyFloatBarrier

\section{Galaxy overdensity in redshift space}

Having shown that the velocity field predicted by Lagrangian perturbation theory adequately traces that in simulations, we now proceed to model the galaxy overdensity in redshift space and compare it against simulations to characterize the quality and error of the model.
\vskip 4pt

\subsection{Model of the galaxy overdensity in redshift space}

A perturbative model for the galaxy overdensity field in redshift space can be derived following the same procedure as for the real space modeling~\cite{Marcel1811}. One of the key ingredients needed for a successful model are large displacements induced by the long-wavelength density fluctuations. If they are not accounted for properly, the model will fail on scales smaller than $\mathcal O(10)\; {\rm Mpc}$, since the positions of the over- or underdensities will be wrong. It is important to stress that this decorrelation with the nonlinear field is much larger than the naive expectation from the one-loop calculation~\cite{Marcel1811,Taruya:2018jtk}. Indeed, while the effects of the bulk flows are guaranteed to cancel in any $n$-point correlation function (in real or redshift space) due to the Equivalence Principle~\cite{Creminelli:2013nua,Creminelli:2013poa,Kehagias:2013yd}, their impact on the level of realizations of density fields is much more dramatic. For this reason it is natural to use Lagrangian perturbation theory as a description for the nonlinear galaxy density field, since it properly captures the effect of large displacements by design. However, since our measurements in simulations are in Eulerian coordinates, it is useful to rewrite the model to resemble the perturbative expansion in Eulerian perturbation theory. We give the details of this derivation in this section.
\vskip 4pt

The galaxy density field realization in Eulerian space, including RSD, can be modeled as follows \cite{Matsubara:2008wx}
\begin{align}
\delta_g^s(\k,\hat\z) = \int d^3 \q\, (1+\delta_g^{\rm L}(\q))  e^{-i\k\cdot\left(\q+\vpsi^s(\q) \right) } \;,
\label{eq:ModelStart}
\end{align}
where the bias expansion in Lagrangian coordinates up to cubic order is given by~\cite{Desjacques:2016bnm}
\begin{align}
\delta_g^{{\rm L}}(\q)\ =\ b_1^{{\rm L}}\,\delta_1(\q)&\,+\,b_2^{{\rm L}}\,[\delta_2(\q)-\sigma_1^2]\, +\,  b_{\G_2}^{{\rm L}}\G_2(\q)\nonumber\\[4pt]
&\, +\,b_3^{{\rm L}}\,\delta_3(\q)\,+\,  b_{\G_2\delta}^{{\rm L}}\,[{\G_2\delta}](\q)\, +\, b_{\G_3}^{{\rm L}}\, \G_3(\q)\,+\, b_{\Gamma_3}^{{\rm L}}\,\Gamma_3(\q)
\; .
\end{align}
In our notation $\delta_n(\q)\equiv \delta_1^n(\q)$ and $\delta_1(\q)$ is the linear density field. The explicit form of all bias operators as well as their relation to other bases used in the literature can be found in~\cite{Desjacques:2016bnm}. We can also use the perturbative expansion of the nonlinear displacement
\be
\vpsi^s(\q) = \sum_{n=1}^\infty R^{[n]} (\hat \z) \cdot \vpsi_n (\q) \; .
\ee 
Since the linear displacement is the largest contribution to $\vpsi$, we can expand all higher order terms from the exponent in \eqq{ModelStart} and treat them as additional nonlinearities in the bias expansion for the galaxy density field. The explicit derivation is given in Appendix~\ref{app:CubicPTmodel}. Here we report only the final result
\begin{align}
\label{eq:main_cubic_model}
\delta_g^s(\k,\hat\z) = \int d^3 \q\, \Big[ 1+ \delta_g^{\rm L} & -\frac 3{14} \G_2 - \frac 3{14} (1+b_1^{\rm L}) \delta_1\G_2 + \frac 16 \Gamma_3 + \frac 19 \G_3  \nonumber \\
& - \frac 37  f \G_2^{\parallel}  - \frac 37  f b_1^{\rm L} \delta_1 \G_2^{\parallel}  - \frac 58  f  \Gamma_3^{\parallel} + \frac 13  f  \G_3^{\parallel}  -  \frac 9{14} f \mathcal K_3  - \frac 3{14}  f^2  \delta_1^{\parallel}  \G_2^{\parallel}   \nonumber \\
& - R^{[2]}_{ij}\psi_2^i \partial_j \big( (1+b_1^L)\delta_1 + f\delta_1^{\parallel} \big)\Big]   e^{-i\k\cdot(\q+R^{[1]} \vpsi_1)}  \;.
\end{align}
Note that the explicit Lagrangian coordinates are suppressed to avoid clutter. The additional nonlinear terms that appear only in redshift space are defined as
\begin{align}
\mathcal O^{\parallel}(\q,\hat \z) & \equiv  \hat z^i \hat z^j \frac{\partial_i\partial_j}{\nabla^2} \mathcal O(\q) \;, \\
\mathcal K_3(\q,\hat \z) & \equiv \hat z_i \hat z_j \frac{\partial_i\partial_m}{\nabla^2} \delta_1(\q) \frac{\partial_m\partial_j}{\nabla^2} \G_2(\q)  \;.
\end{align}
\vskip 4pt

Let us make a couple of comments about this model for the galaxy density field in redshift space. On top of $\delta_g^{\rm L}$, all additional nonlinear operators come from expanding the second and third order displacement from the exponent. Most of these nonlinear terms can be written in the form of bias operators in redshift space, 
with the exception of the second and third line in~\eqref{eq:main_cubic_model}, which represent the second-order shift acting on the linear density field. 
It is worth noting that all new terms have fixed coefficients as expected. 
While this model can appear a bit cumbersome at first sight, it is equivalent (up to two-loop terms that we neglected) to the more familiar results either in Lagrangian~\cite{2013MNRAS.429.1674C,Vlah:2016bcl,Chen:2020fxs} or IR-resummed Eulerian perturbation theory~\cite{Senatore:2014via,Senatore:2014vja,Baldauf:2015xfa,Perko:2016puo,Blas:2016sfa,Ivanov:2018gjr}. Finally, for the purposes of comparing the theory and simulations, all third-order terms in square brackets can be absorbed in a transfer function multiplying $\delta_1$. We can therefore write
\begin{align}
   \delta_g^s(\k,\hat \z) = & \int d^3 \q\, \Big[ 1 - \frac 37  f \G_2^{\parallel}(\vq) \nonumber \\
& + \beta_1(k,\mu)\,\delta_1(\q)\,+\,b_2^{{\rm L}}\,[\delta_2(\q)-\sigma_1^2]\, +\,  \left(b_{\G_2}^{{\rm L}}-\tfrac{3}{14}\right)\G_2(\q)  \Big]   e^{-i\k\cdot(\q+R^{[1]} \vpsi_1(\q))}  \;,
  \label{eq:deltagModelLong}
\end{align}
where $\mu$ is the cosine of the angle between the Fourier mode $k$ and the line-of-sight $\hat \z$: $\mu\equiv\vk\cdot\hat\vz/k$.
We have finally arrived at the point where we can write the simplified model for the galaxy density field in redshift space directly in Eulerian coordinates. Following~\cite{Marcel1811} and defining redshift-space shifted operators as 
\begin{align}
    \tilde{\mathcal{O}}(\vk,\hat \z) = \int d^3 \q\, \mathcal{O}(\vq) e^{-i\k\cdot(\q+R^{[1]} \vpsi_1(\q))}\;,
\end{align}
the model is given by
\begin{align}
\delta_g^s(\k,\hat \z) =\; &
\delta_Z^s(\vk,\hat \z) - \frac{3}{7}f\tilde{\mathcal G}_2^\parallel(\vk,\hat \z) \nonumber \\
& \quad + \beta_1(k,\mu)\tilde\delta_1(\vk,\hat \z) + b_2^{\rm L} \tilde\delta_2^\perp(\vk,\hat \z) 
+ \left(b_{\G_2}^{{\rm L}}-\tfrac{3}{14}\right) \tilde{\mathcal G}_2^\perp(\vk,\hat \z)\;.
\end{align}
Note that the transfer function $\beta_1(k,\mu)$ is defined as
\be
\beta_1(k,\mu) \equiv b_1^{\rm L} + \sum_a c_a \frac{\langle \tilde \delta_1(\k,\hat \z) \tilde {\mathcal O}^{[3]}_a (-\k,\hat\z) \rangle}{\langle \tilde \delta_1(\k,\hat \z) \tilde \delta_1(-\k,\hat \z) \rangle} \;,
\ee
The sum runs over all cubic terms in equation~\eqref{eq:main_cubic_model} and coefficients $c_a$ can be either Lagrangian biases or deterministic constants. Crucially, using the transfer function comes at no price, since the new model has exactly the same power spectrum up to one-loop order as the original galaxy field given in \eqq{main_cubic_model}. Note that this transfer function also contains all higher derivative counterterms and higher derivative bias operators, such as 
\be
\delta_g^{\rm L} (\q) \supset \left( R_1^2 \nabla^2 + R_2^2 (\hat \z \cdot \nabla)^2 + R_3^4 (\hat \z \cdot \nabla)^4 \right) \delta_1(\q) + \cdots \;,
\ee
where $R_i$ are corresponding length scales. Even though the last term seems to be of higher order in perturbation theory, it has been shown that it is very significant, particularly if the fingers of god effect is more pronounced~\cite{Chudaykin:2020hbf}. The contribution of these operators to the transfer function is a simple polynomial in $k$ and $\mu$
\be
\beta_1(k,\mu) \supset R_1^2 k^2 + R_2^2 k^2 \mu^2 + R_3^4 k^4 \mu^4 + \cdots. 
\ee
\vskip 4pt

To obtain the best-possible fit to the data involving second-order fields only, and at the same time test the perturbative model, we can also promote the second-order biases to transfer functions. We will find later that they can be indeed set to constant without affecting the model error much, in agreement with the perturbation theory prediction. With this in mind, the most general model we can write is
\begin{align}
    \label{eq:deltagModel}
\delta_g^s(\k,\hat \z) = \;&
\delta_Z^s(\vk,\hat \z) - \frac{3}{7}f\tilde{\mathcal G}_2^\parallel(\vk,\hat \z) \nonumber \\
& \quad + \beta_1(k,\mu)\tilde\delta_1(\vk,\hat \z) + \beta_2(k,\mu)\tilde\delta_2^\perp(\vk,\hat \z) 
+ \beta_{\mathcal G_2}(k,\mu)\tilde{\mathcal G}_2^\perp(\vk,\hat \z)\;.
\end{align}
The $\beta_n(k,\mu)$ are transfer functions that can absorb a part of the higher order nonlinearities as well as counterterms. The field $\delta_Z^s$ refers to the redshift-space Zel'dovich density,
\begin{align}
    \delta_Z^s(\vk,\hat \z) = \int d^3 \q\,  e^{-i\k\cdot(\q+R^{[1]} \vpsi_1(\q))}\;.
\end{align}
For easier interpretation of the transfer functions, we orthogonalize $\tilde\delta_2$ with respect to $\tilde\delta_1$, and $\tilde{\mathcal G}_2$ with respect to $\tilde\delta_1$ and $\tilde\delta_2$, in every $(k,\mu)$ bin using Gram-Schmidt as in \cite{Marcel1811}.
\vskip 4pt

The model \eq{deltagModel} is rather similar to the real-space model of \cite{Marcel1811}.
The only differences that are not absorbed by transfer functions are the  RSD displacement $R^{[1]}\vpsi_1$ in the exponent and the additional term proportional to $\mathcal G_2^\parallel$. Additionally, transfer functions now depend on $k$ as well as $\mu$. In the same way as in real space \cite{Marcel1811}, we can decide to rewrite the Zel'dovich RSD density $\delta_Z^s$ in terms of the shifted bias operators. For simplicity, however, we will keep the full Zel'dovich density without rewriting it in this way.
We will include the shifted cubic operator $\delta^3(\vq)$ in the model below, because it is simple to add; results are very similar without the cubic term.
\vskip 4pt

\subsection{Evaluating the galaxy overdensity model in a 3D box}

To evaluate the model \eq{deltagModel} in a 3D box we proceed similarly to \cite{Marcel1811}.
We first draw a realization of the linear density in the 3D box.
We then shift the uniform density $1$, the linear Lagrangian-space density $\delta_1(\vq)$, and the second-order fields $\delta^2(\vq)-\sigma^2$, $\mathcal G_2(\vq)$ and $\mathcal G_2^\parallel(\vq)$ by the linear RSD displacement $R^{[1]}\vpsi_1$, and paint the result to a regular grid in Eulerian space.
Given a simulated redshift-space galaxy density $\delta_\text{sim}$, the transfer functions $\beta_n(k,\mu)$ are then computed using ordinary linear least-squares regression in every $(k,\mu)$ bin. This minimizes the squared model error $P_\text{err}(k,\mu)\propto\langle|\delta_\text{sim}(\vk)-\delta_g^s(\vk)|^2\rangle$.\footnote{Having a model with small model error is useful because it has larger signal-to-noise than models with larger model error.
Additionally, it is important to obtain a model for which the parametric form of the model error power spectrum is known so that the total power spectrum, composed of model and noise power, can be predicted; we will test this in the next subsection.}
These transfer functions are smooth functions of $k$ and $\mu$. We will replace them with a simple 7-parameter fit as described at the end of the next subsection and in the appendix.
\vskip 4pt

\subsection{Comparison with simulations}

To test the bias model \eq{deltagModel} for the galaxy density we compare it against the same N-body simulation galaxies described in \secref{VelSimComparison} above, serving as a proxy for SDSS BOSS CMASS galaxies.
We implement RSD by moving galaxies along the line-of-sight according to the subhalo velocity computed with \textsf{Rockstar}.
\vskip 4pt

\begin{figure}[ht]
\centering
\includegraphics[width=0.99\textwidth]{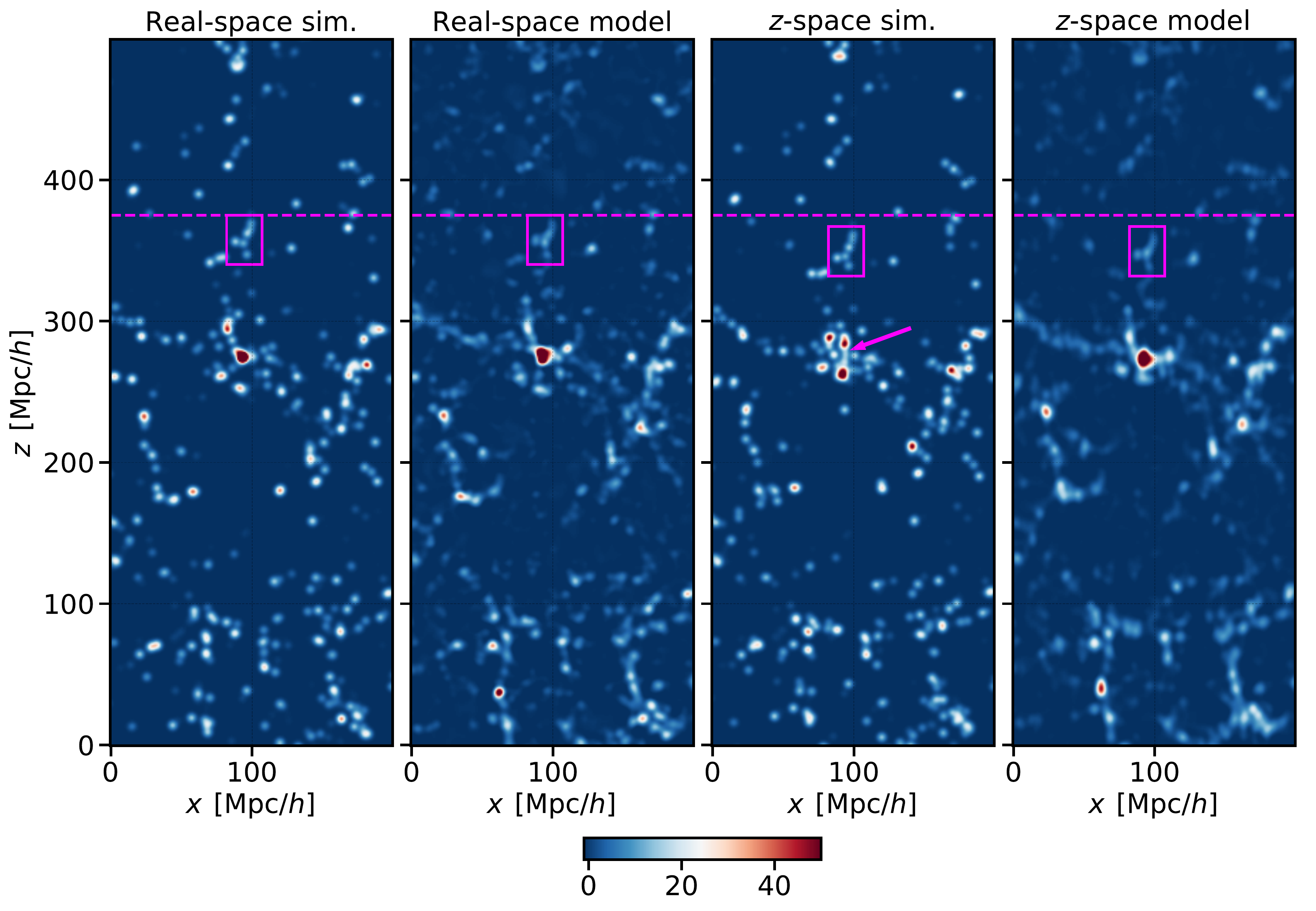}
\caption{Galaxy overdensity $\delta_g$ in a 2D slice around the largest halo in the N-body simulation (red blob in the center; $\log M[h^{-1}\mathrm{M}_\odot]=15.2$). 
From left to right, the panels show the real-space simulation, real-space model, redshift-space simulation and redshift-space model.
In the center there is a Finger-of-God effect, elongating the cluster along the line-of-sight (magenta arrow); this is not captured by the model.
The structure above the cluster (magenta box) moves towards the cluster by about $8\hMpc$; this large-scale flow is captured by the model. 
More typical redshift-space displacements are 3-4$\hMpc$ and difficult to see by eye, but the model matches the simulation well on large scales.
In each panel, the density is smoothed with a $2\hMpc$ 3D Gaussian, and the dimension of each slice is $200\times 1 \times 500\;\hMpc$.
}
\label{fig:deltaSlice}
\end{figure}

\fig{deltaSlice} shows a 2D slice of the resulting simulated galaxy density, the bias model \eq{deltagModel}, and the residual between the two.
The bias model captures the galaxy density well on large scales, but tends to underpredict it in highly overdense regions where the bias model is not applicable.
\vskip 4pt

\begin{figure}[ht]
\centering
\includegraphics[width=0.49\textwidth]{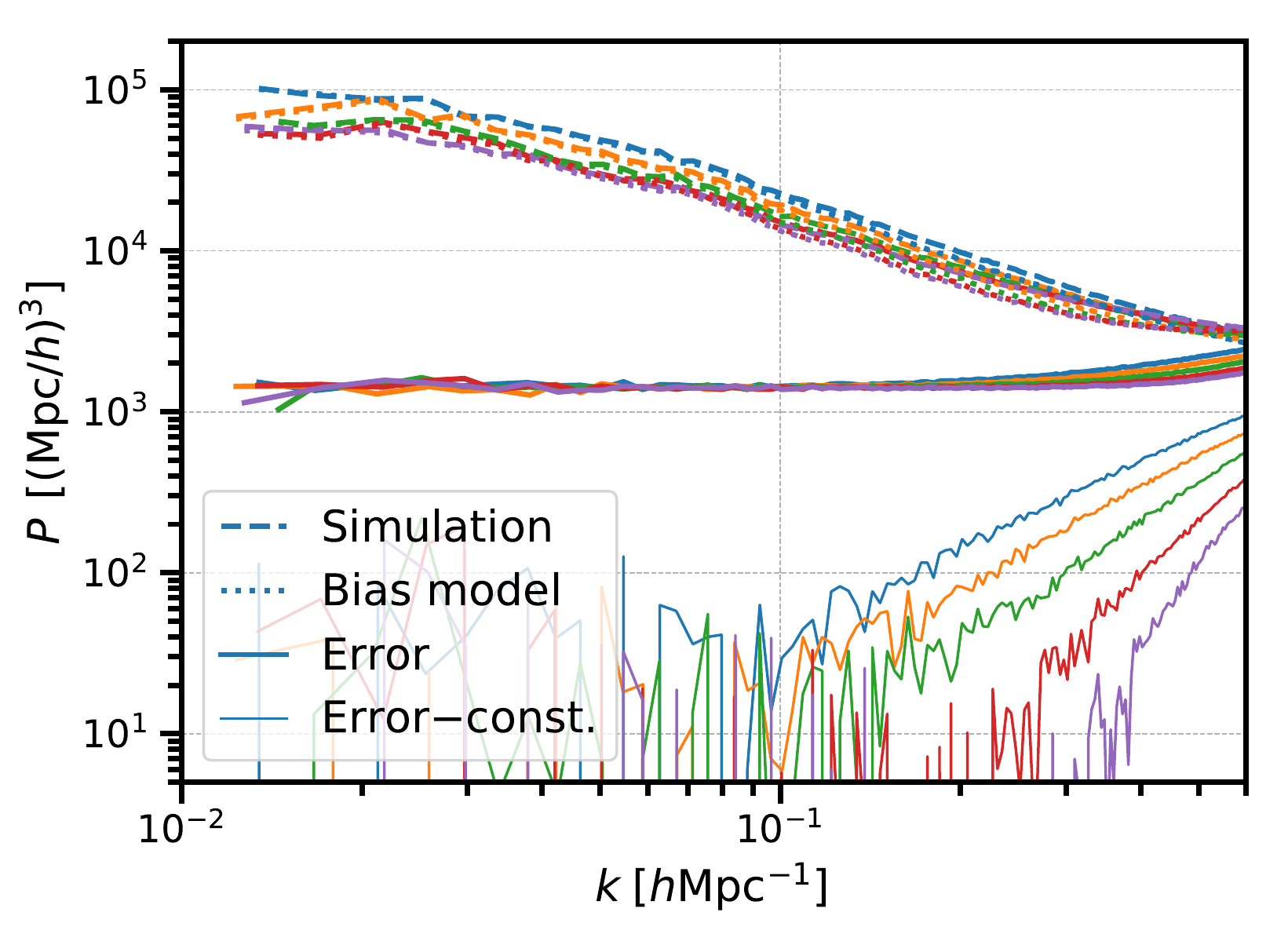}
\includegraphics[width=0.49\textwidth]{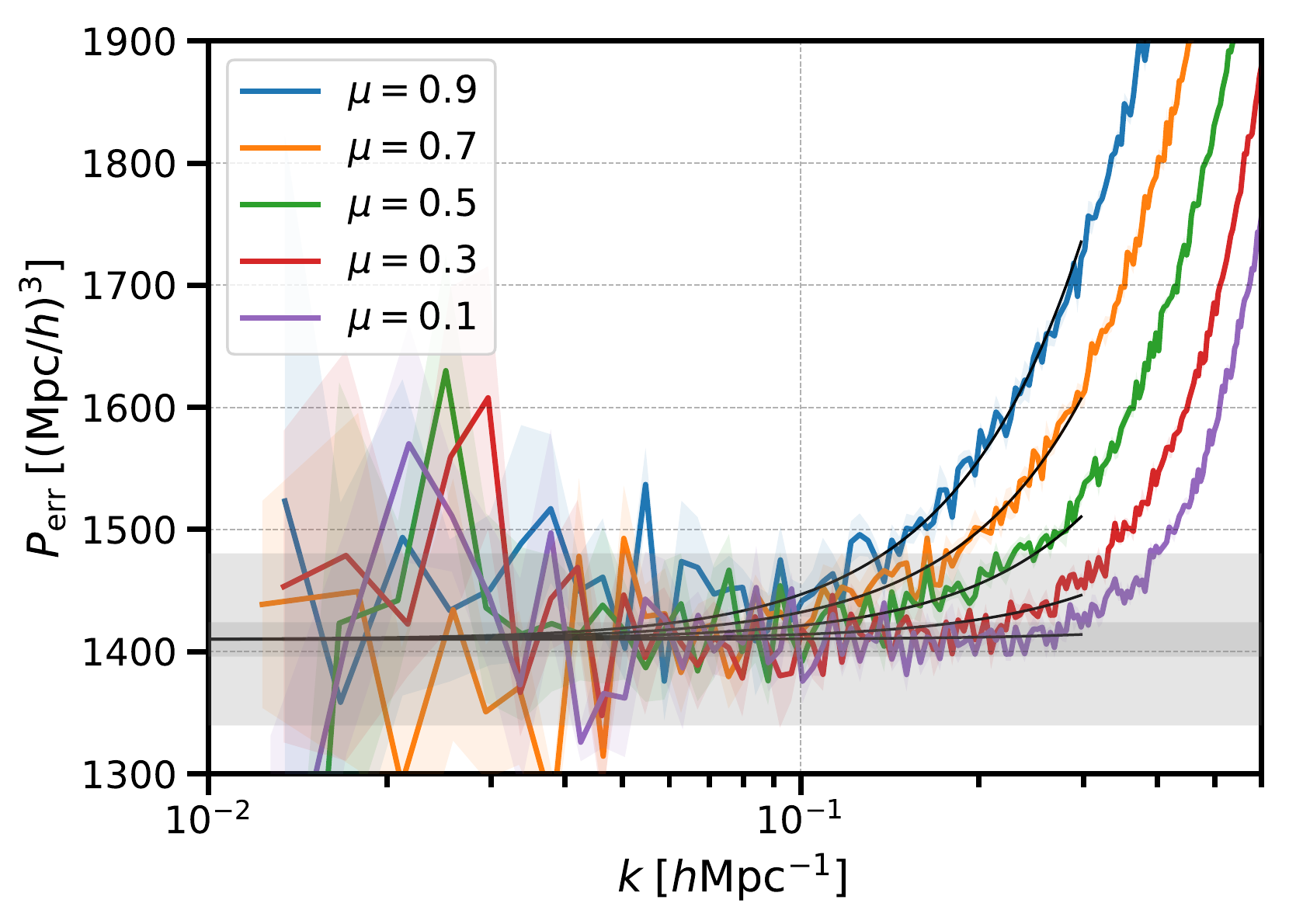}
\caption{\emph{Left panel:} Power spectrum of the simulated galaxy sample (dashed), the bias model (dotted), and the residual error (solid), and error power spectrum minus a constant fit at low $k$ (thin solid).
The power spectra are measured in 5 bins in the cosine $\mu$ with respect to the line of sight, $\mu=0-0.2, 0.2-0.4$, etc (colors).
\emph{Right panel:} Zoom-in of the error power spectrum, $P_\text{err}(k,\mu)\propto\langle|\delta_\text{sim}(\vk)-\delta_g^s(\vk)|^2\rangle$.
This error power spectrum is well fit by \eqq{PerrFit} (black lines).
In both panels, the bias model \eq{deltagModel} uses transfer functions fitted with seven parameters as shown in \fig{BOSSTkFit} below. 
The simulated galaxies are generated by populating \textsf{Rockstar} subhalos in six independent N-body simulations with $1536^3$ DM particles in $L=1500\hMpc$ cubic boxes evolved to redshift $z=0.6$; the subhalos are populated with galaxies to represent SDSS BOSS CMASS galaxies following \cite{Nishimichi:2020tvu}, with a soft lower mass cutoff of $\log_{10}M_\mathrm{min}[h^{-1}\text{M}_\odot]=12.97$.
}
\label{fig:BOSSPerrFittedTk}
\end{figure}

To investigate the model performance more quantitatively, we go to Fourier space and compute the squared model error $P_\text{err}(k,\mu)\propto\langle|\delta_\text{sim}(\vk)-\delta_g^s(\vk)|^2\rangle$.
This is shown in \fig{BOSSPerrFittedTk}.
To include larger scales and reduce scatter in the plots, we increased the volume of the simulation to $L=1500\hMpc$  per side, still using $1536^3$ DM particles, and averaged over six realizations.
The resulting error power spectrum is constant on large scales, and exhibits a $k^2\mu^2$ correction that becomes important at $k\simeq 0.1\ihMpc$.
This is consistent with the stochastic noise power spectrum derived in \cite{Perko:2016puo}.
Indeed, we find that at $k\leq 0.3\ihMpc$ the error power spectrum is well approximated by
\begin{align}
\label{eq:PerrFit}
P_\text{err}(k,\mu) = 
\frac{1}{\bar n_g} \left(c_{\epsilon,1} + c_{\epsilon,3}f
\mu^2\left(\frac{k}{k_\text{M}}\right)^2\right)
\;,
\end{align}
where
\begin{align}
    c_{\epsilon,1} &\;=\; 0.599\;, \nonumber\\
    c_{\epsilon,3} &\;=\; 2.45\, \left(\frac{k_\text{M}}{1\,h\text{Mpc}^{-1}}\right)^2\;.
\end{align}
The number density of simulated galaxies is $\bar n_g=4.25\times 10^{-4}\;h^3\text{Mpc}^{-3}$ and the logarithmic growth rate is $f=0.786$ at redshift $z=0.6$. The amplitude of the noise is compatible with real space results for similar halo number density~\cite{Marcel1811}. The amplitude of the scale dependent part of the noise is related to the stochastic velocity dispersion and it is consistent with measurements in the previous section and values of the counter terms measured in large-volume simulations~\cite{Nishimichi:2020tvu} and real data~\cite{DAmico:2019fhj,Ivanov:2019pdj,Philcox:2020vvt}. 
From \fig{BOSSPerrFittedTk} it is clear that both stochastic noise terms are detected with high significance in the error power spectrum.
This demonstrates that the field-level model for the galaxy density in redshift space is accurate with errors as expected theoretically up to $k\simeq 0.3\ihMpc$, for the galaxies extracted from our simulations. Since agreement at the field level is a much more stringent test than comparing power spectra only, we can see these results as yet another nontrivial check that the one-loop power spectrum in redshift space is indeed the adequate model to describe galaxy clustering on large scales. 
\vskip 4pt

We do not find evidence for an isotropic $k^2$ correction to the error power spectrum, although this term can be present theoretically \cite{Perko:2016puo}. This is consistent with the noise of halos in real space, and can be understood by considering the scale corresponding to the typical halo size \cite{Marcel1811}.
Such a $k^2$ correction may be present for other tracers, especially if they probe larger halos, and on smaller scales.
\vskip 4pt

In \app{BOSSFreeTk} we show the transfer functions obtained by minimizing the error power spectrum in every $(k,\mu)$ bin, and fits of them using a 7-parameter fitting function (4 parameters to describe the scale dependence of $\beta_1$, and 3 parameters to fit the other transfer functions with constants). The error power spectrum shown in \fig{BOSSPerrFittedTk} assumes these fitted transfer functions, as they are more smooth and therefore more realistic (although the error power spectrum for fully free transfer functions looks similar, see \fig{PerrBOSSFreeTk} in the appendix).
Instead of the fitting functions, one could use perturbation theory to predict the functional form of the transfer functions and then fit for the bias parameters, as done in real space in \cite{Marcel1811}; we leave this to future work.
\vskip 4pt

In \app{DESI} we show results when including lower-mass subhalos, that will be observed by DESI.
The error power spectrum is again well described by \eqq{PerrFit}, of course with different values for the fitting parameters.
\vskip 4pt

\MyFloatBarrier
\section{Conclusions}

Previous work modeled the overdensity realization of dark matter particles and dark matter halos in real space \cite{Baldauf:2015zga,Taruya:2018jtk,Marcel1811}. 
Here, we generalized this approach to galaxies in redshift-space.
To model the redshift space distortions caused by the peculiar velocities of galaxies, we shift galaxy bias operators by an additional displacement along the line of sight predicted by the first-order Lagrangian-space velocity.
We then calibrate transfer functions to obtain the best-possible deterministic large-scale model using these shifted galaxy bias operators.
The resulting model captures the redshift-space galaxy overdensity in N-body simulations well on perturbative scales.
\vskip 4pt

We computed the stochastic noise of the model by subtracting the model prediction from the simulated galaxy overdensity.
The power spectrum of this noise is white on large scales, $k\lesssim 0.1\ihMpc$, for the BOSS CMASS-like and the lower-mass mock galaxies we considered at redshift $z=0.6$.
On smaller scales, the noise becomes anisotropic and scale-dependent. It increases along the line of sight and towards smaller scales.
This is expected from noise of galaxy velocities that enters the redshift space distortions along the line of sight.
We find that for mildly nonlinear scales, $k\lesssim 0.3\ihMpc$, the anisotropic and scale-dependent correction to the white noise power spectrum is well fit by a $k^2\mu^2$ term, where $\mu$ is the cosine with the line of sight; see \eqq{PerrFit}.
This parametric form of the stochastic noise power spectrum agrees with the theoretical expectation \cite{Perko:2016puo}.
We do not find evidence for an additional $k^2$ term, but this may be specific to the mock galaxies that we used and may be present for other galaxy samples. 
These results provide new important evidence that the one-loop power spectrum (including all relevant counter terms) with the anisotropic scale-dependent noise is a good model for galaxy clustering in redshift space on large scales and yet another justification for using it in analyzing the real data~\cite{DAmico:2019fhj,Ivanov:2019pdj,Philcox:2020vvt}. 
Furthermore, the field level measurements of the difference between simulations and the model provide the most realistic estimates of the so-called theoretical error, and they can be used instead of templates based on perturbation theory or calibration to simulations at the power spectrum level~\cite{Chudaykin:2020hbf}. 
\vskip 4pt

We also studied the continuous velocity field as predicted by perturbation theory and compared it against simulated galaxy velocities.
We found that roughly 90\% of the simulated galaxies have a velocity that matches the large-scale flows predicted by perturbation theory, while the remaining 10\% of galaxies have larger velocity errors that are responsible for half of the rms velocity error.
Identifying the galaxies with large velocity errors and removing them in observational settings could be beneficial for parameter inference because it could allow modeling smaller scales with a modest increase in shot noise; however it is challenging to identify these galaxies with large velocity errors observationally.
\vskip 4pt

One aspect of our work that could be improved is that we fitted transfer functions that enter the bias model with simple smooth functions in $k$ and $\mu$ rather than modeling their parametric form from first principles. Doing the latter would be an important next step to further check the properties of the noise.
Another limitation is that galaxies in galaxy surveys may have different clustering properties and velocities than the simulated mock galaxies that we considered (subhalos in dark matter-only N-body simulations).
It is also possible to extend the galaxy bias model to higher order in perturbation theory, both in terms of bias operators and in terms of the order used for the displacement field.
\vskip 8pt

\subsection*{Acknowledgements}
It is a pleasure to thank G.~Cabass, E.~Castorina, A.~Moradinezhad Dizgah, T.~Nishimichi, M.~Takada and Z.~Vlah for helpful discussions.
M.S.~thanks IPMU at the University of Tokyo for the hospitality and  useful discussions.
Simulations and numerical analyses were performed on the Helios cluster at IAS and used the public software packages \textsc{MP-Gadget}  \cite{yu_feng_2018_1451799}, \textsc{nbodykit} \cite{Hand:2017pqn} and \textsc{Rockstar} \cite{Rockstar}.
M.S.~acknowledges support from the Corning Glass Works Fellowship and the National Science Foundation. 
M.I.~is partially supported by the Simons Foundation's \textit{Origins of the Universe} program. 
O.P.~acknowledges funding from the WFIRST program through NNG26PJ30C and NNN12AA01C.
\vskip 4pt

\MyFloatBarrier

\bibliographystyle{utphys}
\bibliography{refs}

\MyFloatBarrier

\appendix

\section{Linear Eulerian perturbation theory velocity, and Friend-of-Friends based tracers}
\label{app:OtherVelocityPlots}
\begin{figure}[h]
\centering
\includegraphics[width=0.8\textwidth]{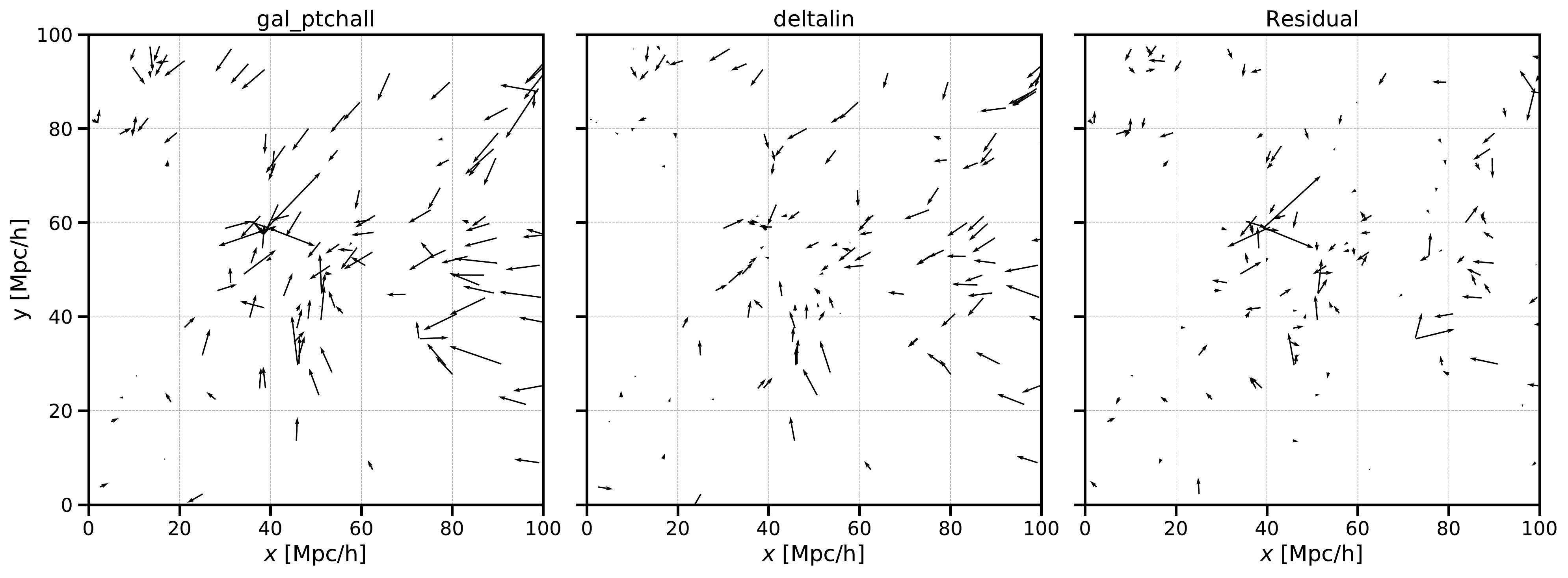}
\caption{Same as \fig{PTChallVel}, but modeling the RSD displacement corresponding to the linear velocity $\vk/k^2 \delta_1(\vk)$.
This is visibly worse than the LPT model used in \fig{PTChallVel}.
}
\label{fig:PTChallVelLin}
\end{figure}

It is instructive to compare the simulated velocities also with other models for the velocity. \fig{PTChallVelLin} shows the prediction of linear Eulerian perturbation theory for the velocity, $\vk/k^2 \delta_1(\vk)$. This is clearly worse than the LPT-based model in the main text, which takes into account the volume distortions according to the Zeldovich approximation. 
\vskip 4pt

\begin{figure}[h]
\centering
\includegraphics[width=0.8\textwidth]{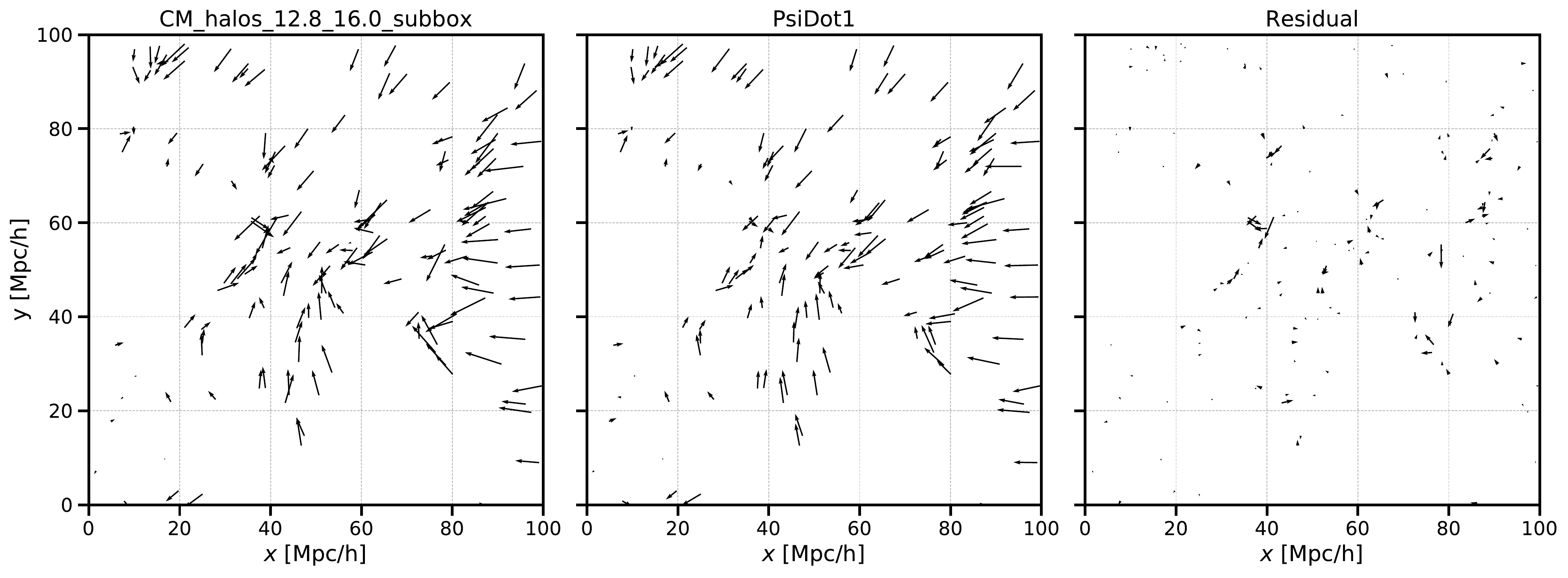}
\caption{Same as previous figure, but for the center-of-mass position and velocity of FOF halos with $\log M[h^{-1}\mathrm{M}_\odot]\ge 12.8$.
The LPT velocity model from the main text works rather well for these center-of-mass FOF halos, without showing the large residuals seen in the main text and in the previous figure. 
This shows that main mistake of the model comes from the fact that galaxy or \textsf{Rockstar} subhalo velocities differ from the center-of-mass velocity of FOF halos. This is caused by the fact that the subhalo velocity can be offset from the halo core velocity, and due to virial motion of subhalos (e.g.~\cite{Rockstar}).
}
\label{fig:CMHalosVel}
\end{figure}

\begin{figure}[h]
\centering
\includegraphics[width=0.8\textwidth]{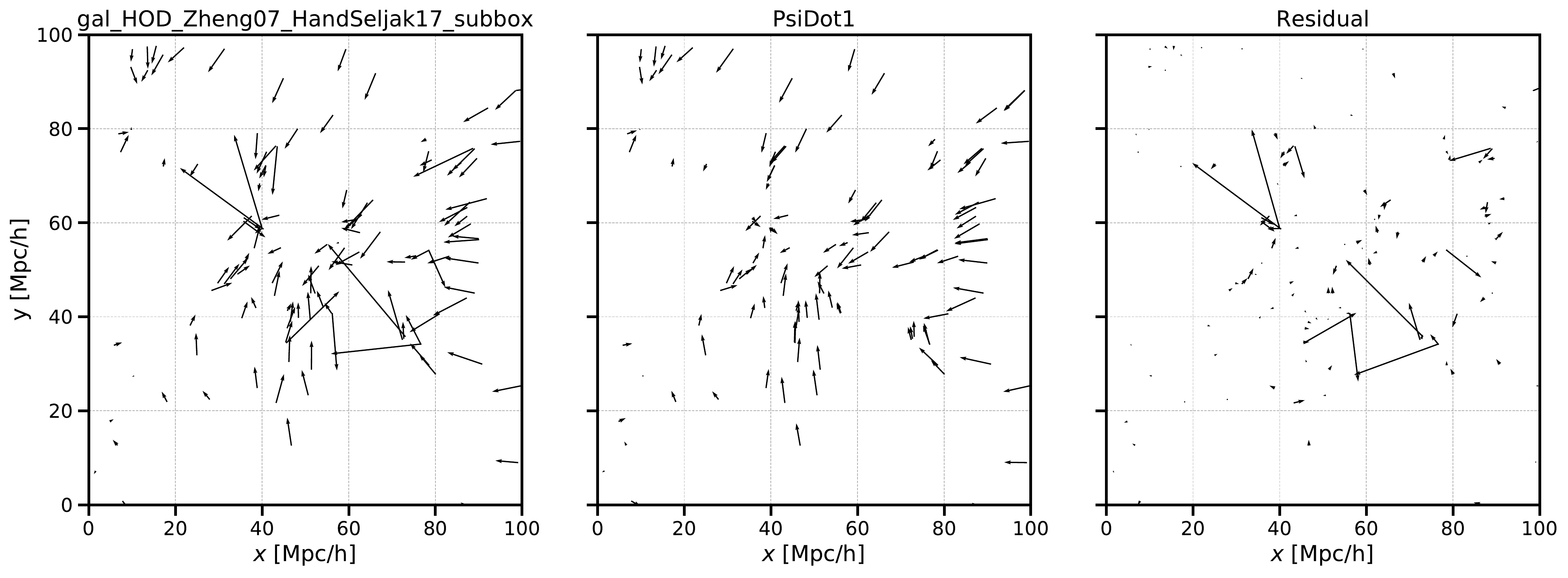}
\caption{Same as previous figure, but for HOD galaxies painted into FOF halos, compared with the velocity model from the main text. This looks similar to the galaxy sample in the main text, though the HOD satellites seem to have larger velocities here. These velocities are based on the NFW profile assigned to FOF halos.
}
\label{fig:HODVel}
\end{figure}

One may also wonder whether objects based on friends-of-friends (FOF) halos exhibit similar velocities.
In \fig{CMHalosVel} we show that the velocity of center-of-mass halos can be modeled much better. This is not surprising, because center-of-mass halos follow the large-scale velocity field and there are less or no virial motions involved.
In \fig{HODVel} we show that HOD galaxies painted on FOF halos can produce velocities that are more similar to the Rockstar galaxies described earlier, though there are notable differences, at least for the HOD parameters chosen here (same as in \cite{Hand:2017ilm}). 
\vskip 4pt

\MyFloatBarrier
\section{Transfer functions for BOSS CMASS-like galaxies}
\label{app:BOSSFreeTk}

\begin{figure}[htbp]
\centering
\includegraphics[width=0.5\textwidth]{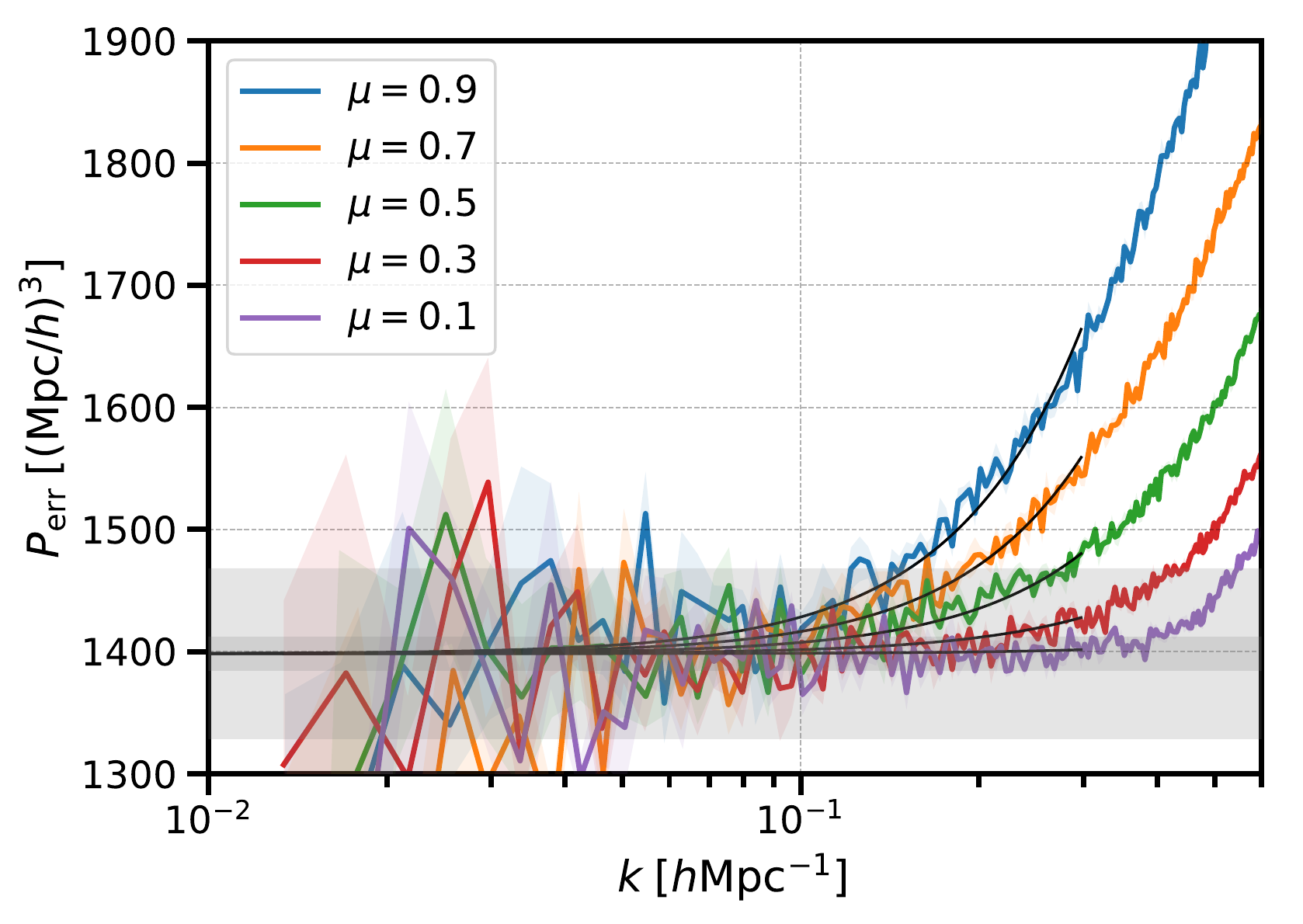}
\caption{Error power spectrum when using transfer functions $\beta_n(k,\mu)$ that are allowed to be free functions of $k$ and $\mu$ and determined by minimizing the error power spectrum. 
}
\label{fig:PerrBOSSFreeTk}
\end{figure}

\begin{figure}[htbp]
\centering
\includegraphics[height=0.48\textheight]{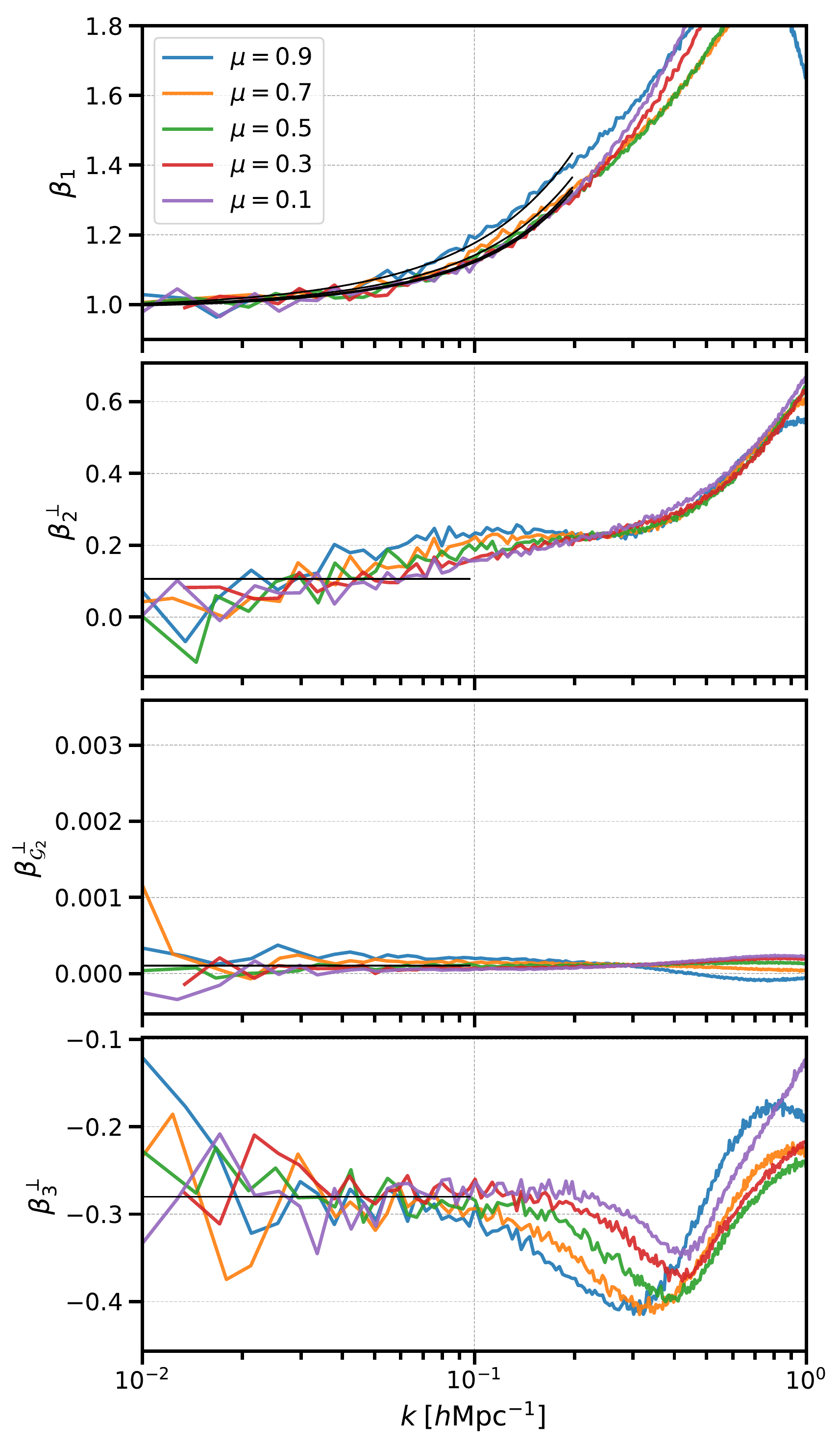}
\caption{\emph{Colored lines:} Transfer functions when modeling the simulated galaxies with the bias model
\eq{deltagModel}.
\emph{Black lines:}
Fit of transfer functions with $\beta_1=0.99 + 0.96 k + 3.9 k^2 + 0.84k\mu^4$ and $(\beta_2^\perp,\beta_{\mathcal{G}_2}^\perp,\beta_3^\perp)=(0.11, 1\times 10^{-4}, -0.28)$, where $k$ is in units of $\ihMpc$. Note that the transfer function $\beta_3$ corresponds to simplest cubic operator $\tilde{\delta}_3$.
}
\label{fig:BOSSTkFit}
\end{figure}

\fig{BOSSTkFit} shows the transfer functions computed by minimizing the model error for the simulated BOSS CMASS-like galaxies.
This can be fitted by a 7-parameter fitting function as described in the caption. 
These fits to the transfer functions (black lines in \fig{BOSSTkFit}) are used to compute the model error in the main text.
If one instead uses the full transfer functions (colored lines in \fig{BOSSTkFit}), the resulting error power is slightly lower, but not by much; see \fig{PerrBOSSFreeTk}.
This shows that the transfer functions can be approximated by smooth functions parameterized by seven numbers, and that the model does not require the hundreds of free parameters corresponding to transfer functions that are free functions of $k$ and $\mu$. 

\MyFloatBarrier
\section{Lower-mass halos}
\label{app:DESI}

\begin{figure}[ht]
\centering
\includegraphics[width=0.49\textwidth]{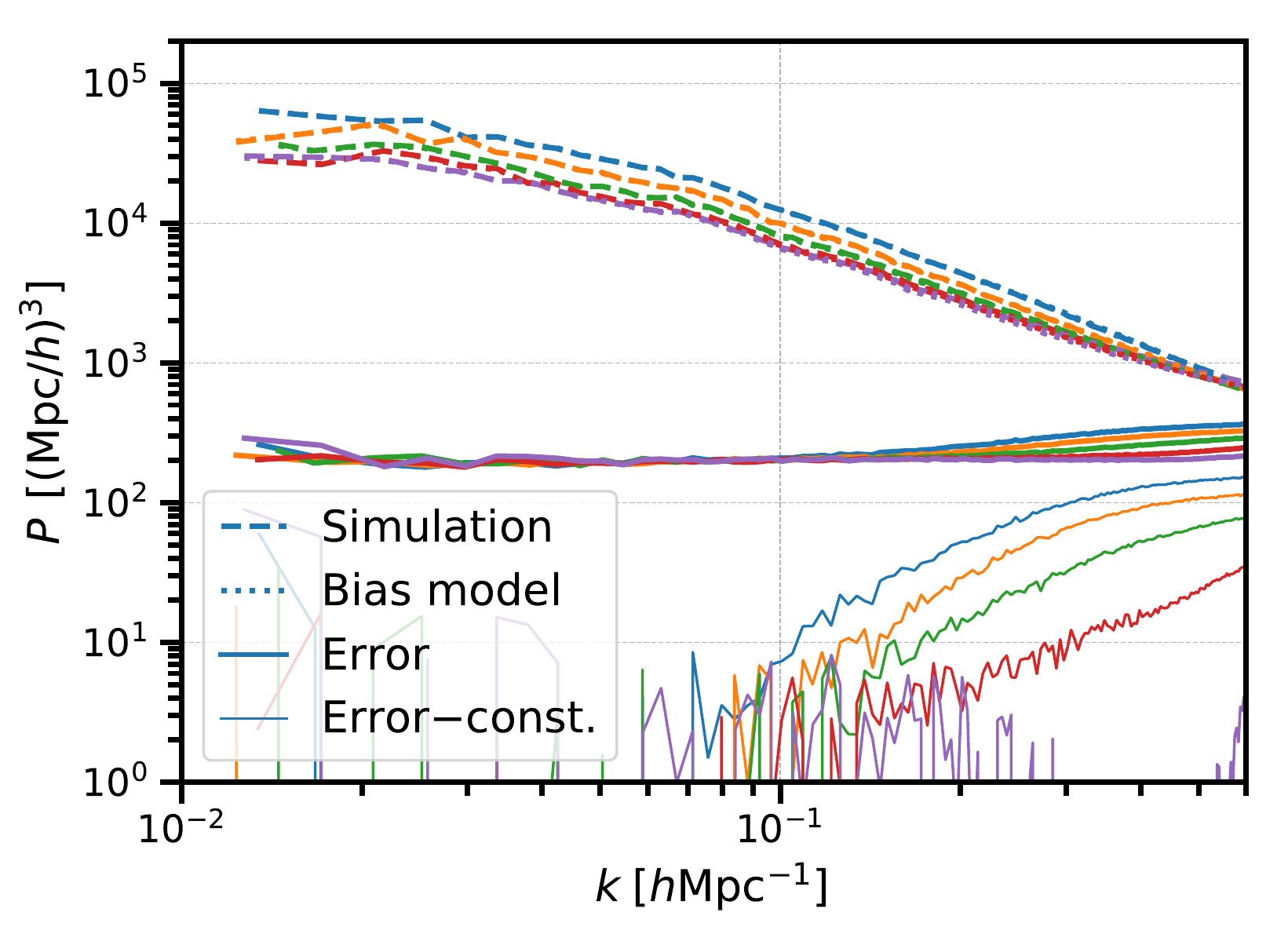}
\includegraphics[width=0.49\textwidth]{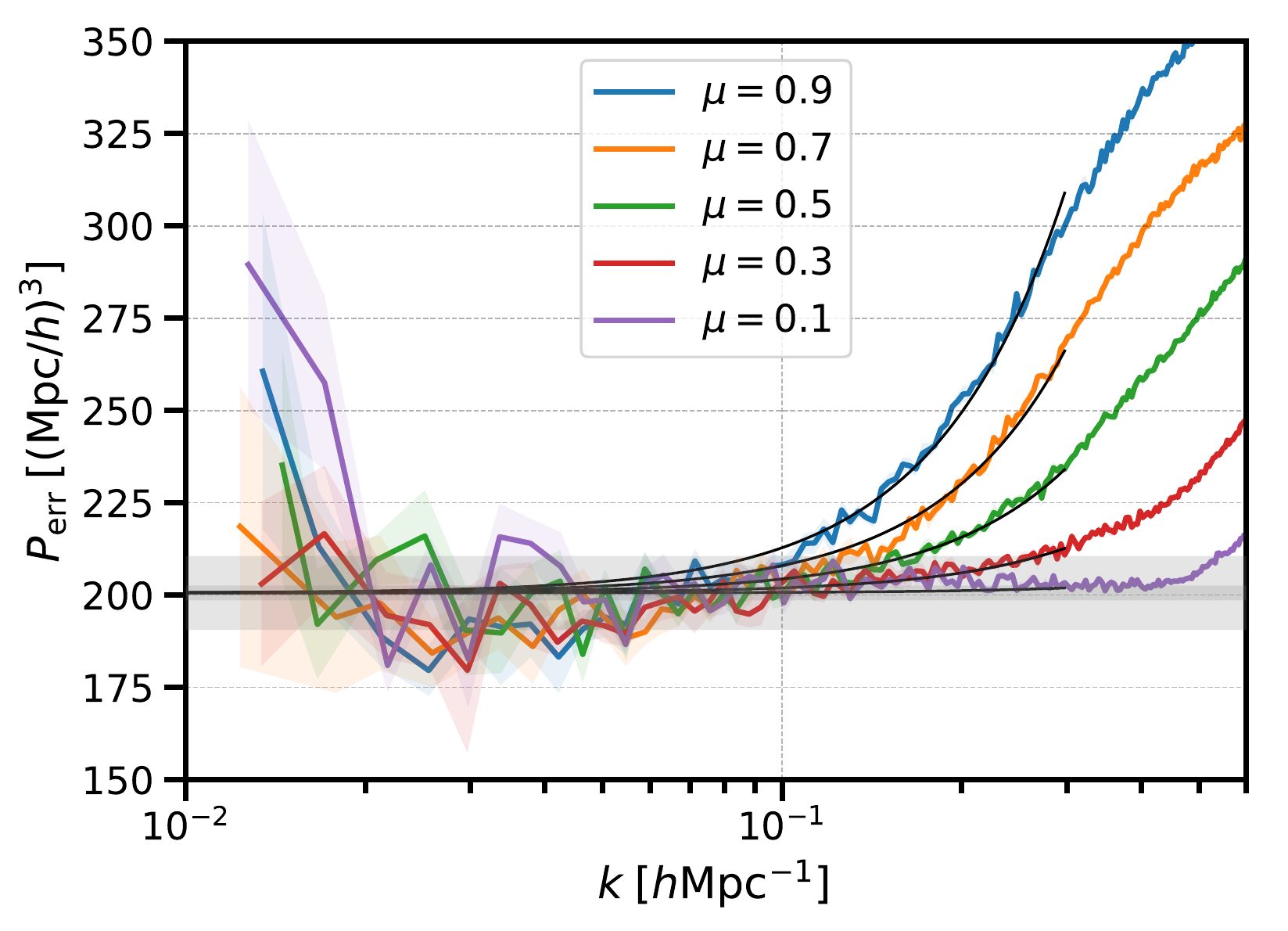}
\caption{Simulation, model, and error power spectrum $P_\text{err}(k,\mu)$ for lower-mass subhalos than in the main text, using a soft lower mass cutoff of $\log_{10}M_\mathrm{min}[h^{-1}\text{M}_\odot]=11.5$ instead of $12.97$. 
The error power spectrum is well described by \eqq{PerrFit} (black lines in the right panel).
The bias model uses transfer functions that are fitted with a 7-parameter fitting formula as shown in \fig{DESITk} as 
resulting from the bias model with transfer functions as shown in the previous plot.
}
\label{fig:DESIPerr}
\end{figure}

\begin{figure}[ht]
\centering
\includegraphics[height=0.48\textheight]{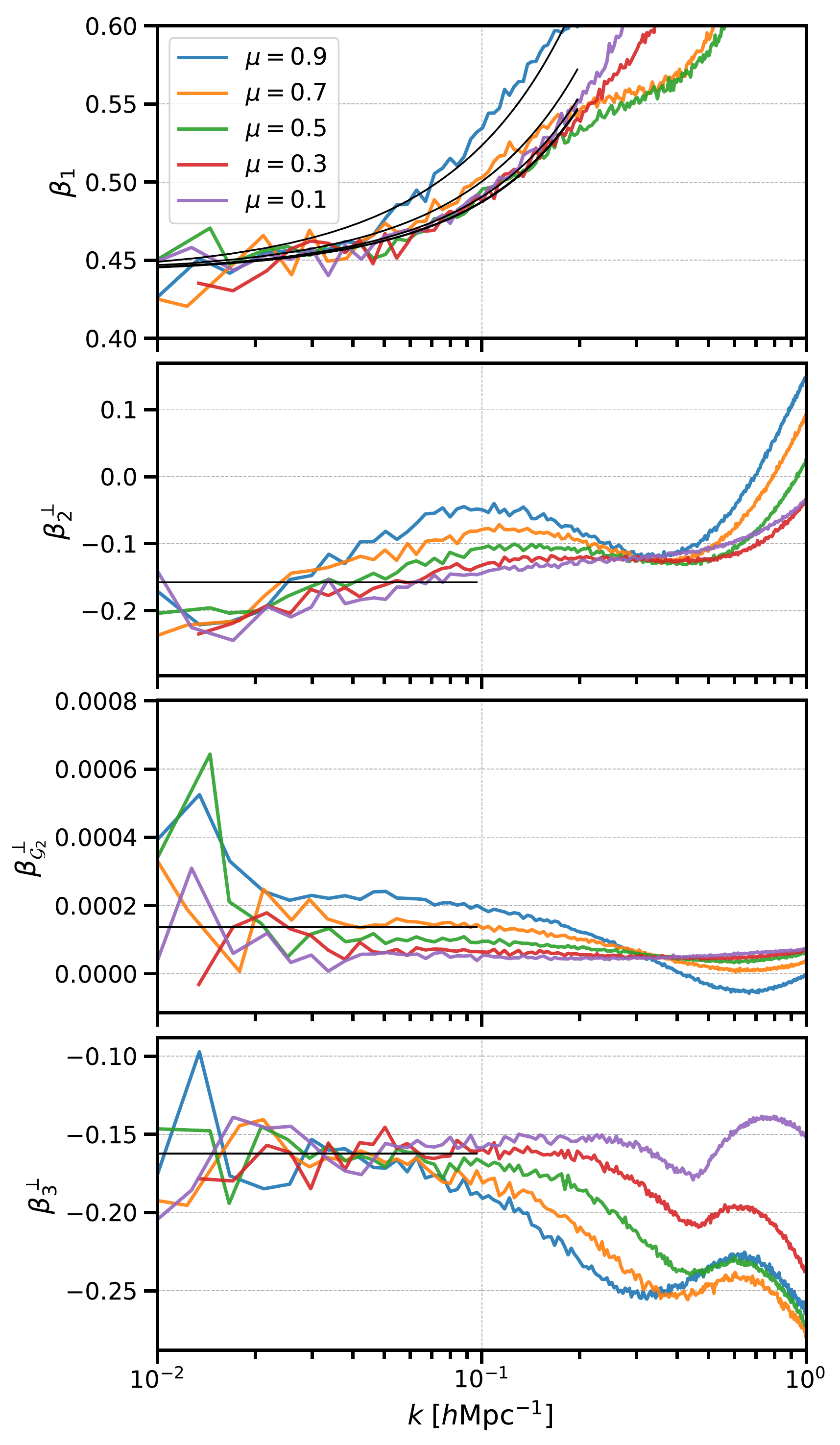}
\caption{Transfer functions for the lower-mass subhalos. Fit of transfer functions: $\beta_1=0.44 + 0.38 k + 0.76 k^2 + 0.55 k\mu^4$ and $(\beta_2^\perp,\beta_{\mathcal{G}_2}^\perp,\beta_3^\perp)=(-0.16, 1\times 10^{-4}, -0.16)$.
}
\label{fig:DESITk}
\end{figure}

In this appendix we repeat the analysis of the main text for a sample that also includes galaxies in lower-mass subhalos.
Specifically, we populate \textsf{Rockstar} subhalos with galaxies by choosing
$\log_{10}M_\mathrm{min}[h^{-1}\text{M}_\odot]=11.5$ for the mass cut-off (instead of $12.97$ used in the main text).
The resulting transfer functions of the bias model is shown in \fig{DESITk}. 
These are again well-described by a 7-parameter fitting function. 
The error power spectrum of the model using these fitted transfer functions is shown in \fig{DESIPerr}.
This error power spectrum is again well described by \eqq{PerrFit} at $k\lesssim 0.3\ihMpc$. The amplitudes of the stochastic noise contributions are now
\begin{align}
    c_{\epsilon,1} &\;=\; 0.723 \;, \nonumber\\
    c_{\epsilon,3} &\;=\; 6.93 \, \left(\frac{k_\text{M}}{h\,\text{Mpc}^{-1}}\right)^2\;,
\end{align}
where the number density of this galaxy sample is $\bar n_g=3.6\times 10^{-3}\;h^3\text{Mpc}^{-3}$.
The $k^2\mu^2$ stochastic term is larger than for the higher-mass sample considered in the main text. 
This is likely caused by the additional satellite galaxies included in the lower-mass sample, with larger velocity dispersion.
Note that the sample is still based on dark matter-only N-body simulations which do not include quenching or other effects related to star formation. This makes it nontrivial to extrapolate the results to galaxies that are selected by specific survey targeting strategies.

\MyFloatBarrier
\section{Galaxy overdensity in redshift space to cubic order in perturbation theory}
\label{app:CubicPTmodel}
In this Appendix we derive the cubic model for the galaxy density field in redshift space used in the main text. We start from 
\begin{align}
\delta_g^s(\k,\hat\z) = \int d^3 \q\, (1+\delta_g^{\rm L}(\q))  e^{-i\k\cdot\left(\q+\vpsi^s(\q) \right) } \;,
\end{align}
and expand the nonlinear displacement terms in the exponent. Omitting the explicit Lagrangian coordinates for simplicity, we get
\begin{align}
\delta_g^s(\k,\hat\z) &= \int d^3 \q\, (1+\delta_g^{\rm L})\Big[ 1-i\k\cdot \bar \vpsi \Big]  e^{-i\k\cdot(\q+ R^{[1]} \vpsi_1)} \nonumber \\
& = \int d^3 \q\,  e^{-i\k R^{[1]} \vpsi_1 } (1+\delta_g^{\rm L}) \Big[ 1+ \bar\vpsi \cdot \nabla_q \Big]  e^{-i\k\cdot\q} \nonumber \\
& = \int d^3 \q\,  e^{-i\k R^{[1]} \vpsi_1 } \Big[ 1+ \delta_g^{\rm L} + ik_i (1+\delta_g^{\rm L}) R^{[1]}_{il} \partial_j \psi_1^l \bar \psi^j - (1+\delta_g^{\rm L}) \nabla \cdot \bar\vpsi - \bar \vpsi \cdot \nabla \delta_g^{\rm L} \Big]  e^{-i\k\cdot\q}  \;,
\end{align}
where
\be
\bar \vpsi (\q,\hat\z) \equiv R^{[2]}(\hat\z) \vpsi_2(\q) + R^{[3]}(\hat\z) \vpsi_3(\q) \;.
\ee
Now we can start simplifying this expression. In the third term in the square brackets we can drop $\delta_g^{\rm L}$ since it multiplies expression that is at least cubic in perturbation theory. We can further replace $k_i$ with a derivative with respect to $q_i$ of the exponent, do one more integration by parts and write
\begin{align}
\delta_g^s(\k,\hat\z) =  \int d^3 \q\, \Big[ 1+ \delta_g^{\rm L} + \bar \psi^j \partial_j (R^{[1]}_{il} \partial_i \psi_1^l) &+ R^{[2]}_{il} R^{[1]}_{jm} \partial_i \psi_1^m \partial_j \psi_2^l 
 - (1+\delta_g^{\rm L}) \nabla \cdot \bar \vpsi \nonumber\\
 &\qquad\qquad\;\; - \bar \vpsi \cdot \nabla \delta_g^{\rm L} \Big]
 e^{-i\k\cdot(\q+R^{[1]} \vpsi_1)}  \;,
\end{align}
keeping only operators up to cubic order as always. We have achieved that the only $k$ dependence is in the exponent, and therefore all the terms in the sum can be represented as some shifted operator. Next, we combine the third and the last term in the square brackets as follows
\be
\bar\psi^j \partial_j (R^{[1]}_{il} \partial_i \psi_1^l) - \bar \vpsi \cdot \nabla \delta_g^{\rm L} = - R^{[2]}_{ij}\psi_2^i \partial_j \big( (1+b_1^{\rm L})\delta_1 + f\delta_1^{\parallel} \big) + {\rm higher\; orders}  \;.
\ee
Note that this operator is the shift of the linear redshift space solution by the second order redshift space displacement. This is exactly the structure we expect at cubic order in perturbation theory, and it is a good check that all relevant terms have been kept in the derivation. Expanding $\bar\vpsi$ everywhere, we find
\begin{align}
\delta_g^s(\k,\hat\z) = \int d^3 \q\, \Big[ 1+ \delta_h^{\rm L} + & R^{[2]}_{il} R^{[1]}_{jm} \partial_i \psi_1^m \partial_j \psi_2^l - (1+b_1^{\rm L} \delta_1) R^{[2]}_{ij}\partial_i\psi_2^j  - R^{[3]}_{ij}\partial_i\psi_3^j    \nonumber \\
& \qquad - R^{[2]}_{ij}\psi_2^i \partial_j \big( (1+b_1^L)\delta_1 + f\delta_1^{\parallel} \big)\Big]   e^{-i\k\cdot(\q+R^{[1]} \vpsi_1)}  \;.
\end{align}

To further simplify this formula we can exploit the fact that all perturbative displacements can be written in terms of the appropriate bias operators as follows~\cite{Marcel1811}
\be
\psi_1^i = -\frac{\partial_i}{\nabla^2} \delta_1\;,  \qquad \quad   \psi_2^i =  \frac 3{14} \frac{\partial_i}{\nabla^2} \G_2  \qquad \quad {\rm and } \qquad \quad  \psi_3^i = \frac{\partial_i}{\nabla^2} \left( \frac 5{24} \Gamma_3 - \frac 19 \G_3 \right) \;.
\ee
Now it is easy to see that the following relations hold 
\begin{align}
R^{[2]}_{ij}\partial_i\psi_2^j  & = \frac 3{14} \G_2 + \frac 37  f \G_2^{\parallel} \;, \\
R^{[3]}_{ij}\partial_i\psi_3^j & =  \frac 5{24} \Gamma_3 - \frac 19 \G_3 + \frac 58  f  \Gamma_3^{\parallel} - \frac 13  f  \G_3^{\parallel} \;, \\
R^{[2]}_{il} R^{[1]}_{jm} \partial_i \psi_1^m \partial_j \psi_2^l & = -\frac 3{14} \delta_1\G_2 + \frac 38 \Gamma_3  -\frac 3{14}  f^2  \delta_1^{\parallel}  \G_2^{\parallel}  -  \frac 9{14} f \mathcal K_3 \;.
\end{align}
Putting everything together, we finally arrive at the desired form of the cubic model used in the main text
\begin{align}
\delta_g^s(\k,\hat\z) = \int d^3 \q\, \Big[ 1+ \delta_g^{\rm L} & -\frac 3{14} \G_2 - \frac 3{14} (1+b_1^{\rm L}) \delta_1\G_2 + \frac 16 \Gamma_3 + \frac 19 \G_3  \nonumber \\
& - \frac 37  f \G_2^{\parallel}  - \frac 37  f b_1^{\rm L} \delta_1 \G_2^{\parallel}  - \frac 58  f  \Gamma_3^{\parallel} + \frac 13  f  \G_3^{\parallel}  -  \frac 9{14} f \mathcal K_3  - \frac 3{14}  f^2  \delta_1^{\parallel}  \G_2^{\parallel}   \nonumber \\
& - R^{[2]}_{ij}\psi_2^i \partial_j \big( (1+b_1^L)\delta_1 + f\delta_1^{\parallel} \big)\Big]   e^{-i\k\cdot(\q+R^{[1]} \vpsi_1)}  \;.
\end{align}

\label{lastpage}

\end{document}